\documentclass[12pt]{article}

\usepackage{amsmath}
\usepackage{graphicx}
\usepackage{eufrak}
\usepackage{cite}

\renewcommand{\theequation}{\arabic{section}.\arabic{equation}}
\newcommand{\mys}[1]{\section{#1} \setcounter{equation}{0}}

\newcommand{\myappendix}{\appendix
   \renewcommand{\theequation}{\Alph{section}.\arabic{equation}}
   \vspace{30pt} \noindent {\Large \bf Appendices}}

\newlength{\dummysp}
\newcommand{\tr}{\mathop{{\hbox{Tr} \, }}\nolimits}

\newcommand{\half}{\frac{1}{2}}

\newcommand{\beq}{\begin{eqnarray}}
\newcommand{\eeq}{\end{eqnarray}}
\newcommand{\nnn}{ \nonumber \\ }
\newcommand{\p}{{\partial}}

\newcommand{\Rbf}{{{\bf R}}}

\newcommand{\chib}{{\bar \chi}}

\newcommand{\e}{{\epsilon}}
\newcommand{\s}{{\sigma}}
\newcommand{\vev}[1]{{\langle #1 \rangle}}
\newcommand{\bigvev}[1]{{\left\langle #1 \right\rangle}}
\newcommand{\ord}[1]{{{\cal O}(#1)}}
\newcommand{\gappeq}{\mathrel{\rlap {\raise.5ex\hbox{$>$}}
{\lower.5ex\hbox{$\sim$}}}}
\newcommand{\lappeq}{\mathrel{\rlap{\raise.5ex\hbox{$<$}}
{\lower.5ex\hbox{$\sim$}}}}
\newcommand{\myref}[1]{(\ref{#1})}

\newcommand{\bfe}[1]{\vspace{4pt} {\bf #1 \hspace{2pt}}}

\newcommand{\ben}{\begin{enumerate}}
\newcommand{\een}{\end{enumerate}}

\newcommand{\sqtw}{\sqrt{2}}
\newcommand{\fourth}{\frac{1}{4}}
\newcommand{\sbar}{{\bar \s}}
\newcommand{\phib}{{\bar \phi}}

\newcommand{\psib}{{\bar \psi}}

\newcommand{\ddd}{\nnn &&}

\newcommand{\bit}{\begin{itemize}}
\newcommand{\eit}{\end{itemize}}
\newcommand{\Lcal}{{\cal L}}

\newcommand{\susy}{supersymmetry}
\newcommand{\susyc}{supersymmetric}

\newcommand{\Ncal}{{\cal N}}

\newcommand{\Ocal}{{\cal O}}

\newcommand{\lambar}{{\bar \lambda}}

\newcommand{\pbar}{{\bar p}}
\newcommand{\qbar}{{\bar q}}
\newcommand{\pdag}{p^\dagger}
\newcommand{\qdag}{q^\dagger}
\newcommand{\pconj}{p^*}
\newcommand{\qconj}{q^*}
\newcommand{\ta}{(t^a)}
\newcommand{\sighat}{{\hat \s}}
\newcommand{\sbarhat}{{\hat \sbar}}
\newcommand{\eps}{\e}
\newcommand{\nn}{\nonumber}
\newcommand{\gfivehat}{\hat{\gamma}_5}
\newcommand{\gfive}{\gamma_5}
\newcommand{\NfourSYM}{$\Ncal=4$ SYM}

\newcommand{\Nfour}{$\Ncal=4$}
\newcommand{\mres}{m_{\text{res}}}

\newcommand{\mrest}{$\mres$}
\newcommand{\DMR}{DESY-M\"unster-Roma}
\newcommand{\cond}{\vev{\lambar \lambda}}
\newcommand{\condt}{$\cond$}

\newcommand{\IRFP}{IR fixed point}
\newcommand{\xibar}{{\bar \xi}}
\newcommand{\Dcal}{{\cal D}}

\def\[{\left [}
\def\]{\right ]}
\def\({\left (}
\def\){\right )}

\def\nott#1{\setbox0=\hbox{$#1$}                % set a box for #1 
   \dimen0=\wd0                                 % and get its size
   \setbox1=\hbox{/} \dimen1=\wd1               % get size of /
   \ifdim\dimen0>\dimen1                        % #1 is bigger
      \rlap{\hbox to \dimen0{\hfil/\hfil}}      % so center / in box
      #1                                        % and print #1
   \else                                        % / is bigger
      \rlap{\hbox to \dimen1{\hfil$#1$\hfil}}   % so center #1
      /                                         % and print /
   \fi}                                         %

\begin{document}

%\markboth{Giedt}{Four-dimensional lattice supersymmetry}

%\catchline{}{}{}{}{}
%
%%%%%%%%%%%%%%%%%%%%%%%%%%%%%%%%%%%%%%%%%%%%%%%%%%%%%%%%%%%%%%%%%%%%

\begin{titlepage}

\renewcommand{\thefootnote}{\fnsymbol{footnote}}

% DATE 

\hfill Mar.~13, 2009

%\hfill yymm:nnnn

\vspace{0.45in}

\begin{center}
{\bf \Large Progress in four-dimensional \\ \vskip 5pt lattice supersymmetry}
\end{center}

\vspace{0.15in}

\begin{center}
{\bf \large Joel Giedt\footnote{{\tt giedtj@rpi.edu}}}
\end{center}

\vspace{0.15in}

\begin{center}
{\it Department of Physics, Applied Physics and Astronomy, \\ \vskip 2pt
Rensselaer Polytechnic Institute, Troy, NY 12180}
\end{center}

\vspace{0.15in}

\begin{abstract}
We are entering an era where a number of
large-scale lattice simulations of four-dimensional
supersymmetric theories are under way.  Moreover,
proposals for how to approach such studies 
continue to progress.  One particular line of
research in this direction is described here.  General
actions for super-QCD, including counterterms required on the lattice,
are given.  We obtain the number of fine-tunings that is
required, once gauge and flavor symmetries are
accounted for, provided Ginsparg-Wilson fermions are
used for the gauginos.  We also review and extend our recent
work on lattice formulations of $\Ncal=4$ super-Yang-Mills and $\Ncal=1$ super-Yang-Mills
that exploit Ginsparg-Wilson fermions.
\end{abstract}

%\keywords{Lattice supersymmetry; lattice gauge theory; Ginsparg-Wilson fermions;
%domain wall fermions}
%\ccode{PACS numbers: }

\end{titlepage}

\renewcommand{\thefootnote}{\arabic{footnote}}
\setcounter{footnote}{0}

\mys{Introduction}
\subsection{Motivations}
\label{moti}
Supersymmetric lattice field theories are
motivated by the desire to obtain nonperturbative information
that cannot be obtained by other means.
Having reviewed the aims of lattice \susy\ recently in \cite{Giedt:2006pd},
here we will offer only a brief reiteration, adding a
few remarks having to do with recent developments.  

\bfe{Definition.}
A classic example of a nonperturbative anomaly is the {\it Witten
anomaly} \cite{Witten:1982fp}.  A more recent example
is \cite{Hayakawa:2006fd}.  It has been argued that
a nonperturbative supersymmetry anomaly exists \cite{Casher:1999mz}.
The lattice approach is a tool to investigate this question.

\bfe{Nonholomorphic quantities.}
In a theory of {\it chiral superfields} $\phi$,
holomorphic quantities $w(\phi)$ are protected by
nonrenormalization theorems.  When combined with symmetries
and ``the power of holomorphy'' \cite{Seiberg:1994bp},
much can be learned about strong dynamics in \susyc\ theories.
Not so for nonholomorphic quantities $k(\phi,\phib)$;
little is known about them in the strongly coupled regime.
The {\it supersymmetry-breaking
soft-terms} that determine spectra and couplings in
\susyc\ extensions to the Standard Model depend on the nonholomorphic
{\it K\"ahler potential;}
see for example \cite{Kaplunovsky:1993rd,Brignole:1993dj}.
In fact, it has recently been realized that
strong hidden sector effects can lead to
significant modifications of the observable 
sector soft terms \cite{Cohen:2006qc}, with
the potential to solve some long-standing
phenomenological problems, such as the $\mu / B_\mu$
problem in gauge mediation models \cite{Roy:2007nz}. 
A strong coupling computational method
is needed in order to say anything definite.
To answer the crucial question in \cite{Roy:2007nz}---the
sign of an anomalous dimension---lattice
artifacts at the level of 10-20\% may be tolerable.
One goal of the research that we report here,
in \S\ref{sqcd}, is to develop lattice super-QCD
as a tool to study this sort of problem.

\bfe{Dynamical supersymmetry breaking.}
New strong gauge theory interactions are
commonly employed to split the superpartners from the observed
Standard Model spectrum.\footnote{Reviews 
include \cite{Nilles:2004zg,Gaillard:2007jr}
for {\it supergravity mediation} of {\it gaugino condensation,} and
\cite{Giudice:1998bp} for {\it gauge mediation.}} 
Strongly coupled messenger sectors and compositeness within the
supersymmetric standard model \cite{Cohen:1996vb,ArkaniHamed:1997fq,
Luty:1998vr,Volkas:1987cd,Strassler:1995ia,Nelson:1996km} 
can provide both economy to the
models and difficult strong coupling questions
of phenomenological importance at scales of a 
few TeV, hence relevant to the Large Hadron
Collider \cite{Gabella:2007cp}.  Moreover, such
models are well-motivated by warped string 
compactifications such as have been explored in \cite{Gherghetta:2006yq},
or the older, perturbative string compactifications on
toroidal orbifolds \cite{Giedt:2000bi,
Giedt:2001zw,Giedt:2003an,Giedt:2003gv} whose phenomenology
has been extensively studied \cite{Giedt:2000es,Gaillard:2000na,
Giedt:2002ns,Giedt:2002kb,
Giedt:2002jp,Gaillard:2003gt,Giedt:2004wd,Giedt:2005vx}.  Advances in 
lattice \susy\ move us toward addressing these questions.
Indeed, it is important to continue exploring \susyc\ models
other than the most popular scenarios such as the
constrained minimal supersymmetric standard model (CMSSM),
although new possibilities in these well-studied models
continue to be uncovered, as in \cite{Ellis:2008mc}.
Some of the most promising models of dynamical \susy-breaking
involve chiral gauge theories.  This is a very
difficult problem that we will not address here,
though there is some interesting recent work on
formulations using Ginsparg-Wilson fermions and
strong Yukawa couplings \cite{Bhattacharya:2006dc,
Giedt:2007qg,Poppitz:2007tu,Poppitz:2008au,Poppitz:2009gt}.

\bfe{Gauge-gravity duality.}
Recently lattice methods have contributed \cite{Catterall:2008yz,
Catterall:2007me,Catterall:2007fp} to the
evolving understanding of the relationship
between supersymmetric gauge theory and theories of quantum gravity;
in particular, string/M-theory.  In fact, the $\alpha'$ and
string loop corrections to certain effective supergravity
descriptions of string theory in nontrivial backgrounds
are supposed to be encoded in corrections to the 't Hooft
limit of the gauge theory.  Obviously, the lattice theory
at finite $N$ and coupling $g$ will capture these effects,
though we must still take the continuum limit---which includes
somehow restoring the \susy\ broken by the lattice regulator.

It is worth mentioning that in the case
of matrix supersymmetric quantum mechanics, a
non-lattice approach has been developed with
considerable success in \cite{Hanada:2007ti} and
subsequent articles.  These authors
fix the gauge and work directly in momentum space with a
sharp cutoff, for this case of $1+0$ dimensions.
Supersymmetry breaking by the regulator is believed
to be especially mild.  Certainly it vanishes as the momentum
cutoff is removed to infinity, since one then obtains
the unregulated theory, which is finite and requires no subtractions.

A background independent, nonperturbative formulation of
superstring theory is not known.  Nevertheless,
the theory is in much better shape due to successes
that address nonpertubative and background dependent
aspects:
M(atrix) theory \cite{Banks:1996vh,Ishibashi:1996xs}, 
the AdS/CFT correspondence \cite{Maldacena:1997re,Gubser:1998bc,Witten:1998qj},
the PP-wave limit \cite{Metsaev:2001bj,Metsaev:2002re,Blau:2001ne,Berenstein:2002jq}, 
etc.  It would be very interesting to study these formulations
through their relation to super-Yang-Mills (SYM). 
The Matrix theory formulations of string/M-theory,
and the AdS/CFT correspondence,
are expressed in terms of quantum theories of
dimensionally reduced SYM.  The vacuum of the
gauge theory is believed to have a gravitational
meaning.  Detailed studies of the SYM vacuum might
provide useful information with a gravitational
interpretation.

\subsection{Challenges}
Dondi and Nicolai \cite{Dondi:1976tx} pointed out long ago
that since the supersymmetry algebra closes on the generator
of infinitesmal spacetime translation, which is explicitly
broken by the discretization, the supersymmetry
algebra invariably must be modified on the lattice.
The obvious option is to have it close on discrete
translations.  Since the Leibnitz rule does
not hold on the lattice, the supersymmetry algebra
will be violated for general polynomials of lattice
fields; interacting supersymmetric theories
will not be invariant with respect to the lattice 
supersymmetry \cite{Fujikawa:2002ic,Giedt:2004qs}.

The non-invariance of the interacting lattice action
is an $\ord{a}$ effect ($a$ is the lattice spacing) 
that disappears if one takes
the continuum limit of the lattice action.  Unfortunately,
in the quantum theory these violations, which correspond
to $\ord{a}$ irrelevant operators that supplement the continuum
action, play off against ultraviolet (UV) divergences
to give infinite violations of \susy\ in the $a\to 0$
limit.  Another way of stating it is this:
non-irrelevant \susy-violating operators allowed
by the symmetries of the lattice action will
be radiatively generated (e.g., a mass term
for scalar partners of the gauge boson in extended
SYM theories); \susy-violating
relative renormalizations of terms already
present in the bare action also will occur (e.g.,
quartic scalar self-couplings with a coefficient
other than $g^2/2$, where $g$ is the gauge coupling). 

In fact, the situation is similar to
the chiral limit for Wilson fermions, where
the bare mass must be fine-tuned in order to
cancel the effects of the $\ord{a}$ suppressed
irrelevant Wilson mass operator.  Fine-tuning of counterterms
to achieve the desired continuum limit is in
principle always possible, but an efficient nonperturbative method
is required if one wishes to address strongly
interacting theories.  This is the sort of
approach advocated in \S\ref{sqcd}-\S\ref{mcrw} of this
review.

A contrasting situation is the one where symmetries of the lattice theory 
overcome the difficulty of \susy-violating renormalizations.  An
example is the domain wall fermion formulation of $\Ncal=1$
SYM discussed in \S\ref{neq1}.  
In that case, lattice chiral symmetry in the 
form of {\it Ginsparg-Wilson} (GW) fermions \cite{Ginsparg:1981bj}
prevents additive renormalization of the gluino mass in the continuum limit,
and hence the only \susy-violating non-irrelevant operator is forbidden in that limit
by setting the bare mass to zero 
\cite{Curci:1986sm,Maru:1997kh,Neuberger:1997bg,
Nishimura:1997vg,Kaplan:1999jn,Campos:1999du,Fleming:2000fa,
Kirchner:1998nk,Donini:1997hh}.

In fact, use of symmetries to prevent
bad renormalization is a much more general 
approach that can be applied in a number of theories
\cite{Kaplan:2002wv,Cohen:2003qw}
including some that are four-dimensional \cite{Catterall:2005fd,Kaplan:2005ta,
Catterall:2007kn}.  Some early detailed studies of these types
theories have appeared recently \cite{Giedt:2003ve,Giedt:2003vy,Giedt:2004tn,Catterall:2008dv,
Kanamori:2009dk,Kanamori:2008yy}.
Many aspects of this approach were discussed in 
recent reviews \cite{Catterall:2005eh,Giedt:2006pd,
Giedt:2007hz}, which contain more extensive references
to this line of research.  We will not
dwell on such formulations.

Since $\Ncal=4$ SYM and super-QCD have scalars, even with
lattice chiral symmetry for the fermions the scalar
masses and couplings (Yukawa and quartic)
will receive divergent non-\susyc\ corrections in the continuum limit and must be
(nonperturbatively) fine-tuned to \susyc\ values.  For this reason
these theories have always seemed impractical by the fine-tuning approach.\footnote{Other
approaches to $\Ncal=4$ SYM include Refs.~\cite{Catterall:2005fd,Kaplan:2005ta,
Catterall:2007kn},
which involve {\it orbifold} or {\it twisted \susy} lattices.}

We have argued in \cite{n4gw} that for $\Ncal=4$ SYM this opinion is
overly pessimistic.  In this review we extend our line of reasoning to 
super-QCD (SQCD) where
the situation is, confessedly, more challenging.\footnote{Other
approaches to SQCD have been explored 
in \cite{Endres:2006ic,Giedt:2006dd}, albeit
in a two-dimensional context.}  First, however, let
us concentrate on the reasons why $\Ncal=4$ is more practical that
might be naively concluded:
\ben
\item
If one uses GW fermions the gluinos can be kept massless.
\item
The $SU(4)_R$ symmetry can be preserved, 
restricting renormalizations. 
\item
The parameters that must be tuned consist of:
   \bit
   \item
   one scalar mass, 
   \item
   two/four quartic couplings ($N_c=2,3$ vs.~$N_c>3$), and
   \item
   one Yukawa coupling.
   \eit
\een
Next we summarize aspects of the fine-tuning procedure that
help to make nonperturbative adjustment of four/six
parameters ``practical.''

The Yukawa coupling can be tuned by
rescaling the scalar kinetic term.
This is obvious because the Yukawa coupling strength $y$ can always be
absorbed into a redefinition of the scalar fields $\phi \to \phi/y$,
    causing it to reappear in the scalar kinetic term.
Thus, all tunings can be done by
adjusting bosonic terms in the action.  This allows the tunings to be
done by the ``Ferrenberg-Swendsen method''~\cite{Falcioni:1982cz, 
Ferrenberg:1988yz,Ferrenberg:1989ui}, 
exploring a wide swath of coupling constant space ``offline'' from the results
of a single Monte-Carlo simulation. The parameter range available with
good statistics can be enlarged using {\it multicanonical} techniques
\cite{Binder,Baumann:1986iq,Berg:1991cf,kajantie,deForcrand:2002vx,
deForcrand:2002vs}.  Thus we arrive
at the encouraging result that all fine-tuning can be performed through an
offline analysis; i.e., new simulations that require
large numbers of fermion matrix inversions during
molecular dynamics trajectories are not required.

The message is this:  one need only generate
a set of configurations that coarsely cover the parameter space
in the vicinity of the fine-tuned lattice parameters.
 In practice, this neighborhood of the $\Ncal=4$ SYM or SQCD point in
  parameter space would be determined by starting on very small
   lattices, and in fact perturbative calculations would give
   us a good idea where to begin for weak bare couplings on
   such lattices.  This is because on a small lattice there
   is not much separation between the UV and infrared (IR), and hence
   the effective coupling remains weak at the IR scale.  Simulations
   can then be used to move into stronger coupling regimes, so
   that one bootstraps upon previous results in order to
   stay in the \susyc\ window for bare couplings.
Modest computational resources would be able to perform all
the offline fine-tunings, and to carefully study the location of
the supersymmetric point in the bare lattice parameter space.

Similar statements hold for $\Ncal=2$ SYM theory, though we
do not go into details here.  However, we note that fine-tuning with
Wilson fermions has been analyzed by Montvay in \cite{Montvay:1994ze}.
By contrast, if GW symmetry is exploited, the $SU(2)_R$ symmetry
of the continuum can be preserved, which reduces the number
of counterterms significantly.

\subsection{Overview}
We begin in \S\ref{sqcd} with some new results,
providing the most general continuum Lagrangians
for SQCD consistent with the symmetries that the
lattice will preserve.  This then allows us to
write down the lattice actions including all
counterterms that must be included in order to
fine-tune to the \susyc\ point.  We enumerate the
number of fine-tunings in each case, and show
that they can be accomplished through entirely
bosonic reweighting, with one exception.  In \S\ref{neq4} we review
the results of our previous study of similar approach
in $\Ncal=4$ SYM, extending some of the
discussion of how tuning can be implemented.  In \S\ref{mcrw} we give a discussion
of the multicanonical reweighting method, and
explore how it can be implemented for the theory
described in \S\ref{neq4}.  In \S\ref{neq1}
we describe recent large-scale simulations of $\Ncal=1$ SYM
with a domain wall fermion implementation, adding
some results on nonlinear chiral extrapolations.  The
domain wall fermion implementation of $\Ncal=1$ SYM is
in the spirit of \S\ref{sqcd}-\S\ref{neq4}, except
that the GW symmetry is in this case powerful enough
to prevent any counterterms that would have to be fine-tuned.
Conclusions and an Appendix follow these main sections.

\mys{Lattice super-QCD}
\label{sqcd}
Super-QCD extends QCD by adding a fermionic
partner for the gluon and scalar partners for the
quarks.  Since the theories that we are interested
in here are new gauge interactions that are strong
at scales of a TeV or greater, it is only related
to QCD by way of analogy.  The gauge groups that
we consider here will be SU(N), and the number of
flavors will be $N_f$.  The continuum theory is
briefly reviewed in Appendix \ref{sqcth}.

\subsection{SU(2) gauge theory}
The gauge sector consists of the gauge boson
$A_\mu$ and the gaugino $\lambda$.  The gauge action
will be formulated using a massless GW fermion $\lambda$
and the Wilson plaquette action or
some improved version of it.  For numerical stability,
it will be necessary to simulate at nonzero gaugino
mass and then extrapolate to chiral limit $m_\lambda \to 0$.

The matter sector of the theory consists of
$P_I, Q_I$, $I=1,\ldots,N_f$ {\it chiral superfields,}
each containing a complex scalar and a left-handed Weyl
fermion.
Here we distinguish them based on $U(1)_V$ charge, $+1$
for $P$, $-1$ for $Q$, a symmetry of the continuum theory
that will be preserved exactly in the lattice formulation.
On the other hand, $P_I$ and $Q_I$ are both fundamentals
of SU(2), transforming identically.
We denote scalar and left-handed Weyl fermion components 
by $p_I, q_I$ and $\chi_{pI},\chi_{qI}$ respectively.

\subsubsection{General form of invariants}
The scalar quadratic SU(2) invariants are given in Table \ref{bisu2}.
We will impose two other constraints that are
symmetries of the continuum Lagrangian, given
in the Appendix, Eq.~\ref{csqcd}.  The first is CP conservation,
and hence reality of the coefficients.  The second
is a $Z_2$ exchange symmetry that we will call $S$:
\beq
S:  \quad p_I \leftrightarrow q_I, \quad 
\pconj_I \leftrightarrow \qconj_I, \quad \chi_{pI} \leftrightarrow
\chi_{qI}, \quad \chib_{pI} \leftrightarrow
\chib_{qI}.
\label{pqex}
\eeq
Then the most general mass term for the scalars,
suppressing flavor indices, is:
\beq
 m^2 \( \pdag p + \qdag q \).
\label{dumb}
\eeq
The other $SU(2) \times U(1)_V$ invariant
mass term\footnote{Here and below $T$ denotes transpose, while
$\e$ is the two-dimensional Levi-Cevita
tensor with convention $\e_{12}=1=-\e_{21}$.}
\beq
m'^2 (p^T \e q + \pdag \e \qconj)
\label{genm}
\eeq
is ruled out by the $S$ exchange symmetry \myref{pqex}.
Later, for number of flavors $N_f>1$,
we will impose $SU(N_f)_p \times SU(N_f)_q$ flavor symmetry constraints,
and specify the corresponding matrix structure of $m^2$.
This flavor symmetry also forbids \myref{genm}.

\begin{table}
\begin{center}
\begin{tabular}{|cc|cc|} \hline
SU(2) invariant & $U(1)_V$ charge & SU(2) invariant & $U(1)_V$ charge \\ \hline
$\pdag p$ & 0 & $\qdag q$ & 0 \\
$p^T \e q$ & 0 & $\pdag \e \qconj$ & 0 \\ \hline
$\qdag p$ & 2 & $\pdag q$ & -2 \\
$p^T \e p$ & 2 & $\pdag \e \pconj$ & -2 \\
$\qdag \e \qconj$ & 2 & $q^T \e q$ & -2 \\
\hline
\end{tabular}
\caption{Bilinear SU(2) invariants, flavor indices suppressed.
\label{bisu2}}
\end{center}
\end{table}

The independent quartic invariants, 
only taking into account SU(2) gauge invariance and $U(1)_V$
at this point, are built from the bilinear invariants in the
Table~\ref{bisu2}.  Combining the $U(1)_V$ neutral bilinears,
and imposing $CP$ and $S$ symmetries, we arrive at
the ``(0,0)'' quartic Lagrangian:
\beq
\Lcal_{(0,0)} &=& \lambda_1 [ (\pdag p)^2 + (\qdag q)^2 ]
+ \lambda_2 \pdag p \qdag q + \lambda_3 |p \e q|^2
+ \lambda_4 [ (p \e q)^2 + (\pdag \e \qdag)^2] .
\label{00op}
\eeq
The other (0,0) term that can be obtained from
the quadratics in the table is
\beq
 \lambda_5 (\pdag p + \qdag q) (p \e q + \pconj \e \qconj),
\eeq
but it is ruled out by the S symmetry \myref{pqex}.
Likewise we combine the $U(1)_V$ charged bilinears
to obtain the ``(2,-2)'' quartic Lagrangian:
\beq
\Lcal_{(2,-2)} &=& \nu_1 |\qdag p|^2 + \nu_2 ( |p^T \e p|^2 + |q^T \e q|^2 ) \nnn
&& + \nu_3 [ (p^T \e p) (q^T \e q) + (\pdag \e \pconj) (\qdag \e \qconj) ] \nnn
&& + \nu_4 [(\qdag p) (q^T \e q + \pdag \e \pconj) + 
(\pdag q) (\qdag \e \qconj + p^T \e p) ].
\label{2m2op}
\eeq
In fact, the $\nu_4$ term will violate the nonabelian flavor
symmetry $SU(N_f)_p \times SU(N_f)_q$  
and we will end up discarding it for all but
the one flavor case.

Finally, note that we have eliminated
other SU(2) invariant quartic 
operators through relations such as
\beq
(\pdag_I \s^a p_J) (\pdag_K \s^a p_L) =
- 2 (\pdag_I \e \pconj_K) (p_J \e p_L)
\label{fier}
\eeq
using the Fiertz identity
\beq
\sum_{a=1}^3 \s^a_{\alpha \gamma} \s^a_{\beta \delta} =
- 2 \e_{\alpha \beta} \e_{\gamma \delta} + \delta_{\alpha \gamma} 
\delta_{\beta \delta} .
\eeq
In Eq.~\myref{fier}, $I,J,K,L$ are flavor indices.
Identities like \myref{fier} will relate the general Lagrangian
that we are writing down to the \susyc\
theory, since in the latter the quartic interactions
are typically expressed in the form of the left-hand side.
We will return to this below, once flavor symmetry
constraints have been taken into account.
For now we merely state that the $D$-term Lagrangian in the \susyc\ theory is:
\beq
\Lcal_D = -\frac{g^2}{2} \sum_a D^a D^a, \quad
 D^a =  \sum_I \[  \pdag_I \s^a p_I + \qdag_I \s^a q_I \] .
\label{sus4}
\eeq

On comparing \myref{sus4} to the quartic interactions
of the general theory, \myref{00op} and \myref{2m2op},
we see that we have many more quartic interaction 
parameters than in the continuum theory, where there
is only one type of term with a strength determined by
the gauge coupling.  It will be seen below
that this is a general feature of the SQCD theories:  many
fine-tunings are needed due to a large number of quartic
couplings that are allowed.  We postpone the precise
count of finely-tuned parameters until we take into account flavor structures.
We will do that shortly, but first we complete our general
parameterization by considering the Yukawa couplings.

Here it is not hard to check that the SU(2) gauginos
$\lambda^a$, $a=1,2,3$ give rise to the following unique
CP and S symmetric, hermitian Yukawa interactions, written in two-component
notation:  
\beq
\Lcal_y &=& y_1 [ \pdag \lambda \chi_p
+ \chib_p^T \lambar p  + 
\qdag \lambda \chi_q
+ \chib_q^T \lambar q ] \nnn
&& + y_2 [ p^T \e \lambda \chi_q - \chib_q^T \lambar \e \pbar 
+ q^T \e \lambda \chi_p - \chib_p^T \lambar \e \qbar ]
\label{su2y}
\eeq
where 
\beq
\lambda = \lambda^a t^a, \quad \lambar = \lambar^a t^a,
\quad t^a = \half \s^a
\label{soso}
\eeq
and $\s^a$ are Pauli matrices.
It is obvious that this is hermitian.  CP
acts on the fields according to:
\beq
&& p \to \pconj, \quad \chi_p \to \chib_p, \quad
q \to \qconj, \quad \chi_q \to \chib_q, \nnn
&& \lambda \to \lambda^*= \lambar^a t^{aT} = \lambar^T, \quad
\lambar \to \lambda^T.
\eeq
It can be checked that \myref{su2y} is invariant
under this symmetry.  In doing this one must keep
in mind the rules of two-component spinors as
it relates to Grassmann fields, such as 
\beq
\lambar^a \chib_p \equiv
\lambar^a_{\dot\alpha} \chib_p^{\dot\alpha} =
- \chib_p^{\dot\alpha} \lambar^a_{\dot\alpha} 
= \chib_{p \dot\alpha} \lambar^{a \dot \alpha} 
\equiv \chib_{p} \lambar^{a}, 
\eeq
see Appendix B of \cite{Wess:1992cp}.
In the \susyc\ target theory, $y_2=0$ and $y_1=\sqtw g$,
where $g$ is the gauge coupling, together with
the field redefinition $\lambda \to i \lambda, \; \lambar \to -i \lambar$.
See Appendix \S\ref{sqcth} for
further details.  

The $y_2$ terms are forbidden
if there is more than one flavor, which is quite useful
since directly tuning fermionic interaction 
terms is most likely not practical.
Two approaches to the tuning of the $y_1$ term will be
discussed below, both of which involve tuning bosonic
terms relative to this fermionic term.

\subsubsection{One flavor}
The flavor symmetry is simple to analyze
when $N_f=1$, since there are no
flavor indices to add to the operators
that we have just written down.  All 
but the first of the (2,-2) operators
in \myref{2m2op} vanish identically, eliminating $\nu_{2,3,4}$
from consideration.  All of the (0,0) operators 
in \myref{00op} are allowed, and the mass term \myref{dumb}
and both types of Yukawa terms \myref{su2y} are also unrestricted. 
Altogether {\em nine} parameters must be fine-tuned:
\bit
\item
one scalar mass $m^2$,
\item
six quartic couplings $\lambda_{1,2,3,4,5}, \; \nu_1$, and
\item
two Yukawa couplings $y_{1,2}$.
\eit
As mentioned already, in the $N_f=1$ theory two ``fermionic'' parameters
that must be fine-tuned, $y_{1,2}$.  In a simulation
they are buried in the fermionic determinant, as far as the
Boltzmann weight in configuration space is concerned.  Unlike
the bosonic parameters, they cannot be adjusted ``offline'' by
reweighting techniques.  If we appeal to the method advocated in
\cite{n4gw}, then one of the Yukawa fine-tunings can be
done equivalently through introducing a $p,q$ scalar field
strength $Z_{\phi}$ in front of the kinetic term:
\beq
Z_{\phi} ( |Dp|^2+|Dq|^2 ) .
\eeq
This rescaling of the scalars at tree level then provides
a lever to adjust $y_1$.  (Of course, the
mass must also be rescaled $m^2 \to m'^2 \approx Z_\phi m^2$,
and then more precisely tuned, 
to keep the theory near the desired physical mass.)
What is needed, therefore, is
a lattice symmetry that enforces $y_2=0$.  Unfortunately,
none seems to be available.  The chiral symmetry
\beq
p \to e^{i\theta} p, \quad q \to e^{i\theta} q, \quad
\chi_p \to e^{i\theta} \chi_p, \quad \chi_q \to e^{i\theta} \chi_q
\eeq
is anomalous and is of no help.  We conclude that the $N_f=1$
SU(2) SQCD will be very difficult to study in the current
formulation, due to the additional type of Yukawa coupling.

\subsubsection{Two flavors}
The target theory has
a $SU(2)_p \times SU(2)_q$ flavor symmetry
that we will preserve in the lattice action.  The
unique mass term arising from \myref{genm} is
\beq
m^2 \sum_{I=1}^{N_f} \[ \pdag_I p_I +  \qdag_I q_I \],
\label{fftr}
\eeq
with $N_f=2$.  The quartic terms in \myref{00op} allow for six
flavor symmetric terms 
\beq
&& \lambda_1^{(1)} \[ (\pdag_I p_I)(\pdag_J p_J) +  (p \to q) \]
+ \lambda_1^{(2)} \[ (\pdag_I p_J)(\pdag_J p_I) +  (p \to q) \]
\ddd
+ \lambda_1^{(3)} \e_{IJ} \e_{KL} \[ (\pdag_I p_K) ( \pdag_J p_L)
+ (p \to q) \] + \lambda_2 (\pdag_I p_I)(\qdag_J q_J)
\ddd + \lambda_3 (\pdag_I \e \qconj_J) (p_I^T \e q_J)
+ \lambda_4 \e_{IJ} \e_{KL} \[ (p_I^T \e q_K) ( p_J^T \e q_L)
+ \text{c.c.} \], ~~~~~
\eeq
where c.c.~denotes complex conjugate.
Thus it is only $\lambda_1$ that proliferates---into
three parameters---once flavor symmetric combinations
are enumerated.
In \myref{00op} one has the flavor specifications
\beq
&& \nu_1 (\pdag_I q_J) (\qdag_J p_I) +
\nu_2^{(1)} \[ (p_I^T \e p_J) (\pdag_I \e \pconj_J) + (p \to q) \]
\ddd + ~ \nu_2^{(2)} \e_{IJ} \e_{KL} \[ (p_I^T \e p_J) (\pdag_K \e \pconj_L)
+ (p \to q) \] \ddd
+ ~ \nu_3 \e_{IJ} \e_{KL} [ (p_I^T \e p_J) (q_K^T \e q_L) 
+ (\pdag_I \e \pconj_J) (\qdag_K \e \qconj_L) ],
\eeq
with a slight proliferation  $\nu_2 \to \nu_2^{(1)}, \nu_2^{(2)} $ that is
compensated by the fact that
$\nu_4 \equiv 0$ since that type of term always has a lone
$p$ or a lone $q$.  In fact,
this forbids the $\nu_4$ term for all cases $N_f>1$.  
The $y_2$ Yukawa term in \myref{su2y} is forbidden for the same reason
as the $\nu_4$ potential term.  The flavor symmetric
Yukawa term is just ($y_1 \to y$):
\beq
\Lcal_y = y [ \pdag_I \lambda \chi_{pI} 
+ \chib_{Ip}^T \lambar p_I  
+ \qdag_I \lambda \chi_{qI} 
+ \chib_{qI}^T \lambar q_I ] .
\label{yukF}
\eeq
Altogether one has in the $N_f=2$ case the following {\em twelve} tunings to
perform:
\bit
\item
one scalar mass  $m^2$,
\item 
ten quartic couplings  $\lambda_1^{(1)}, \lambda_1^{(2)}, \lambda_1^{(3)}, \lambda_2,
\lambda_3, \lambda_4, \nu_1, \nu_2^{(1)}, \nu_2^{(2)},  \nu_3$, and
\item
one Yukawa coupling $y$.
\eit

As we will describe in more detail below, two approaches can be
taken towards tuning $y$.  In the first case, as
was discussed above and advocated in \cite{n4gw}, one adjusts
the field strength of the scalars $p,q$ in order
to accomplish the same thing as fine-tuning $y$.
In the second case, and this is a new approach
that we propose for the first time here, one takes $y$ to implicitly define the
gauge coupling of the lattice theory and fine-tunes
the Wilson gauge action coefficient $\beta=4/g^2$
until \susy\ is achieved.  The disadvantage of this
second method is 
 that one does not know {\it a priori} what
the bare gauge coupling of the theory is really is!
That is, it is determined in the process of offline
reweighting.  Yet since in a typical application all
one really wants is say three values of $\beta$
with sufficiently fine lattice spacing $a$, in order
to make a continuum extrapolation, it reasonable
to think that selecting three values of $y$
will accomplish the same goal.  In particular, one
ought to bootstrap from small lattices where the
actual value of $\beta$ for the \susy\ theory can
be determined (in the reweighting process) cheaply.

\subsubsection{$N_f > 2$}
The scalar mass terms are given by \myref{fftr}.
The quartic terms in \myref{00op} allow for seven
flavor symmetric terms
\beq
&& \lambda_1^{(1)} \[ (\pdag_I p_I) (\pdag_J p_J) +  (p \to q) \]
+ \lambda_1^{(2)} \[ (\pdag_I p_J) (\pdag_J p_I) +  (p \to q) \]
\ddd
+ \lambda_1^{(3)} \[ t_{IJ}^a t_{KL}^a (\pdag_I p_J)( \pdag_K p_L)
+  (p \to q) \]
+ \lambda_1^{(4)} \[ t_{IJ}^a t_{KL}^a (\pdag_I p_L)( \pdag_K p_J)
+ (p \to q) \] 
\ddd
+ \lambda_2 (\pdag_I p_I)(\qdag_J q_J)
+ \lambda_3^{(1)} (\pdag_I \e \qconj_J) (p_I^T \e q_J) 
+ \lambda_3^{(2)} t_{IJ}^a t_{KL}^a ( \pdag_I \e \pconj_K) ( p^T_J \e p_L ).
\eeq
Here $t_{IJ}^a$ are generators of the flavor group $SU(N_f)$.
In \myref{00op} one has the flavor specifications
\beq
&& \nu_1 (\pdag_I q_J) (\qdag_J p_I) +
\nu_2^{(1)} \[ (p_I^T \e p_J) (\pdag_I \e \pconj_J) + (p \to q) \]
\ddd 
+ ~ \nu_2^{(2)} \[ t_{IJ}^a t_{KL}^a (\pdag_I \e \pconj_K) (p_J^T \e p_L) 
+ (p \to q) \] .
\eeq
The Yukawa couplings are given by \myref{yukF}.  Altogether
one has {\em twelve} parameters to fine-tune:
\bit
\item
one scalar mass  $m^2$,
\item 
ten quartic couplings  $\lambda_1^{(1)}, \ldots, \lambda_1^{(4)}, 
\lambda_2, \lambda_3^{(1)}, \lambda_3^{(2)}, \nu_1, \nu_2^{(1)}, \nu_2^{(2)}$, and
\item
one Yukawa coupling $y$.
\eit

\subsection{SU(3) gauge theory}
Here triality (the $Z_3$ center symmetry
of $SU(3)$ gauge theory) is rather 
restrictive when combined with the other
symmetries.  The mass term is just as in the SU(2) theory above,
Eq.~\myref{dumb}.
Cubic potential terms are ruled out by $U(1)_V$.  The quartic
potential terms must be built from SU(3)
singlet combinations of one of the forms:\footnote{
Here we follow the convention of raised indices on
the $\bar 3$ irreducible representation ({\it irrep}), 
and hence find it convenient to
write $\pbar$ rather than $\pconj$}
\beq
\pbar^i p_j \pbar^k p_\ell, \quad
\pbar^i p_j \qbar^k q_\ell, \quad
\qbar^i q_j \qbar^k q_\ell.
\eeq
because of $U(1)_V$ and triality; here $ijkl$ are color indices (=1,2,3).
To form color invariants we begin with enumerating the irreducible
representations that occur from pairing:
\beq
3 \times 3 = \bar 3 + 6, \quad 3 \times \bar 3 = 1 + 8,
\eeq
which in terms of fields takes the forms given in the Table \ref{pai3}.
Some words of clarification are in order.  First, conventional
shorthands such as
\beq
\pdag p \equiv \pbar^i p_i , \quad
\pdag t^a p \equiv \pbar^i \ta^{j}_{i} p_j
\eeq
and
\beq
(pq)^i= \half \e^{ijk} p_j q_k, \quad
(pq)_{ij} = \half \( p_i q_j  + p_j q_i\)
\eeq
have been employed.  Here, $\ta^j_i = (1/2) \lambda^a_{ij}$
with $\lambda^a$ the Gell-Mann matrices.
Second, flavor indices have been
suppressed, but are necessary to render the compact
notation sensible.  For example,
expressions such as $\e^{ijk} p_j p_k$
require $N_f>1$ in order to have nonvanishing
result, $\e^{ijk} p_{j,J} p_{k,K}$, $J,K=1,\ldots,N_f$.
Expressions such as $p_i p_j  + p_j p_i$ would
have a flavor specification $p_{i,I} p_{j,J}  + p_{j,I} p_{i,J}$.
For brevity, in Table \ref{pai3} we have left out representations that
can be obtained by complex conjugation, such
as the $3$ representation $\e_{ijk} \pbar^j \pbar^k$.

\begin{table}
\begin{center}
\begin{tabular}{|c|c|c|c|} \hline
quadratic & irrep & $Q_V$ & $Z_3$ \\ \hline
$\pdag p$, ~~ $\qdag q$ & 1 & 0 & 0 \\
$\qdag p$ & 1 & 2 & 0 \\ \hline
$(pq)^i$ & $\bar 3$ &  0 & 2 \\ 
$(pp)^i$ & $\bar 3$ &  2 & 2 \\ 
$(qq)^i$ & $\bar 3$ & -2 & 2 \\ \hline 
$(pq)_{ij}$ & 6 & 0 & 2 \\
$(pp)_{ij}$ & 6 & 2 & 2 \\
$(qq)_{ij}$ & 6 & -2 & 2 \\ \hline
$\pdag t^a p$, ~~ $\qdag t^a q$ & 8 & 0 & 0 \\ 
$\qdag t^a p$ & 8 & 2 & 0 \\ 
$\pdag t^a q$ & 8 & -2 & 0 \\ 
\hline
\end{tabular}
\caption{SU(3) representations from pairs, flavor indices suppressed.
\label{pai3}}
\end{center}
\end{table}

To obtain singlets we  take the combinations
which may be schematically denoted
$1 \cdot 1$, $\bar 3^i 3_i$, $\bar 6^{ij} 6_{ij}$, $8^a 8^a$.  
However two constraints  relate these:
\beq
&& (\pbar \qbar)_i (pq)^i = \fourth \[ (\pdag p) (\qdag q) - (\qdag p)(\pdag q) \],
\ddd
(\pbar \qbar)^{ij} (pq)_{ij} = \half \[ (\pdag p) (\qdag q) + (\qdag p)(\pdag q) \] .
\label{hts}
\eeq
Thus the $\bar 3^i 3_i$ and $\bar 6^{ij} 6_{ij}$ singlets can
be eliminated in favor of the $1 \cdot 1$ forms,
and one finds that the most general quartic Lagrangian is:
\beq
&& \Lcal_4 = \lambda_1 [ (\pdag p)^2 + (\qdag q)^2 ]
+ \lambda_2 (\pdag p) (\qdag q) 
+ \lambda_3 (\pdag q) (\qdag p)
\ddd + \lambda_4 [ (\pdag t^a p)^2 + (\qdag t^a q)^2 ]
+ \lambda_5 (\pdag t^a p)(\qdag t^a q)
+ \lambda_6 (\pdag t^a q)(\qdag t^a p),
\label{su3lam}
\eeq
where as usual flavor specifications remain
to be given (below), depending on the value of $N_f$.
Finally, the Yukawa couplings are
\beq
\Lcal_y = y[ \pbar^T \lambda \chi_p
+ \chib_p^T \lambar p  + 
\qbar^T \lambda \chi_q
+ \chib_q^T \lambar q ] ,
\label{su3y}
\eeq
where $\lambda = \lambda^a t^a$ and $\lambar = \lambar^a t^a$,
as in Eq.~\myref{soso}.  Note that the second type
of term appearing in \myref{su2y} is not allowed,
due to SU(3) triality.

\subsubsection{$N_f=1$}
Here there is nothing to specify; the expressions just
given suffice, with \myref{dumb} for the most
general mass term, Eq.~\myref{su3y} for the
Yukawa terms and \myref{su3lam} for the
quartic terms.  Altogether we have {\em eight} fine-tunings:
\bit
\item one scalar mass $m^2$,
\item six quartic couplings $\lambda_{1,2,\ldots,6}$, and
\item one Yukawa coupling $y$.
\eit
As before, the tuning of the Yukawa can be effectively
accomplished either through the bare scalar field strength $Z_\phi$
or through tuning the bare coupling $\beta=6/g^2$.

\subsubsection{$N_f=2$}
A few operators proliferate because
of different ways of realizing the $SU(2)_p \times SU(2)_q$ flavor
symmetry.  The quartic terms are: 
\beq
\Lcal_4 = &&
\lambda_1^{(1)} \[ (\pdag_I p_I)(\pdag_J p_J) +  (p \to q) \]
+ \lambda_1^{(2)} \[ (\pdag_I p_J)(\pdag_J p_I) +  (p \to q) \]
\ddd
+ \lambda_1^{(3)} \e_{IJ} \e_{KL} \[ (\pdag_I p_K) ( \pdag_J p_L) + (p \to q) \]
+ \lambda_2 (\pdag_I p_I)(\qdag_J q_J)
\ddd
+ \lambda_3 (\pdag_I q_J) (\qdag_J p_I)
+ \lambda_4^{(1)} [ (\pdag_I t^a p_I)(\pdag_J t^a p_J) + (p \to q) ]
\ddd
+ \lambda_4^{(2)} [ (\pdag_I t^a p_J)(\pdag_J t^a p_I) + (p \to q) ]
+ \lambda_4^{(3)} \e_{IJ} \e_{KL} [ (\pdag_I t^a p_K)(\pdag_J t^a p_L) + (p \to q) ]
\ddd
+ \lambda_5 (\pdag_I t^a p_I)(\qdag_J t^a q_J)
+ \lambda_6 (\pdag_I t^a q_J)(\qdag_J t^a p_I).
\eeq
The mass terms are given by \myref{fftr} and the Yukawa terms by \myref{yukF}.
Altogether we have {\em twelve} fine-tunings:
\bit
\item one scalar mass $m^2$,
\item ten quartic couplings $\lambda_{2,3,5,6}$, $\lambda_{1,4}^{(1,2,3)}$, and
\item one Yukawa coupling $y$.
\eit

\subsubsection{$N_f>2$}
We have in this case twelve operators in the quartic Lagrangian:
\beq
\Lcal_4 = &&
 \lambda_1^{(1)} \[ (\pdag_I p_I) (\pdag_J p_J) +  (p \to q) \]
+ \lambda_1^{(2)} \[ (\pdag_I p_J) (\pdag_J p_I) +  (p \to q) \]
\ddd
+ \lambda_1^{(3)} \[ t_{IJ}^a t_{KL}^a (\pdag_I p_J)( \pdag_K p_L)
+  (p \to q) \]
+ \lambda_1^{(4)} \[ t_{IJ}^a t_{KL}^a (\pdag_I p_L)( \pdag_K p_J)
+ (p \to q) \] 
\ddd
+ \lambda_2 (\pdag_I p_I)(\qdag_J q_J)
+ \lambda_3 (\pdag_I q_J) (\qdag_J p_I)
+ \lambda_4^{(1)} [ (\pdag_I t^a p_I)(\pdag_J t^a p_J) + (p \to q) ]
\ddd
+ \lambda_4^{(2)} [ (\pdag_I t^a p_J)(\pdag_J t^a p_I) + (p \to q) ]
+ \lambda_4^{(3)} [ t_{IJ}^a t_{KL}^a (\pdag_I t^a p_J)(\pdag_K t^a p_L) + (p \to q) ]
\ddd
+ \lambda_4^{(4)} [ t_{IJ}^a t_{KL}^a (\pdag_I t^a p_L)(\pdag_K t^a p_J) + (p \to q) ] 
+ \lambda_5 (\pdag_I t^a p_I)(\qdag_J t^a q_J)
\ddd
+ \lambda_6 (\pdag_I t^a q_J)(\qdag_J t^a p_I)
\eeq
The mass terms are given by \myref{fftr}  and the Yukawa terms by \myref{yukF}.
We now have {\em fourteen} fine-tunings:
\bit
\item one scalar mass $m^2$,
\item twelve quartic couplings $\lambda_1^{(1)} , \ldots, \lambda_1^{(4)} ,
\lambda_2, \lambda_3, \lambda_4^{(1)}, \ldots, \lambda_4^{(4)}, \lambda_5, 
\lambda_6$, and
\item one Yukawa coupling $y$.
\eit
Clearly the task of tuning these parameters such
that the long distance effective theory is the much
simpler Lagrangian \myref{csqcd} poses an enormous
challenge.  A first task is to design a strategy for
confirming from lattice data that the effective
potential reduces to Eq.~\myref{sus4}.  We leave
this as a topic for future research.  

\subsection{SU(4) gauge theory}
Here an analysis similar to what has just been performed
for SU(3) leads to the conclusion that the only new quartic
operator, not contained in \myref{su3lam}, is
\beq
\e^{ijk\ell} p_i p_j q_k q_\ell + \text{c.c.}
\label{neop}
\eeq
In particular, using $4 \times 4 = 6 + 10$ to form
quadratics in the $6$ and $10$ representations,
one finds identities similar to \myref{hts} for
$6^{ij} 6_{ij}$ and $10_{ij} \overline{10}^{ij}$
that reduce these to $1 \cdot 1$ forms.
Thus the general quartic Lagrangian is:
\beq
&& \Lcal_4 = \lambda_1 [ (\pdag p)^2 + (\qdag q)^2 ]
+ \lambda_2 (\pdag p) (\qdag q) 
+ \lambda_3 (\pdag q) (\qdag p)
\ddd + \lambda_4 [ (\pdag t^a p)^2 + (\qdag t^a q)^2 ]
+ \lambda_5 (\pdag t^a p)(\qdag t^a q)
+ \lambda_6 (\pdag t^a q)(\qdag t^a p)
\ddd
+ \lambda_7 [ \e^{ijk\ell} p_i p_j q_k q_\ell + \text{c.c.} ] .
\label{su4lam}
\eeq
The mass and Yukawa terms are the same as for SU(3).

The flavor specifications also follow SU(3).  
For $N_f=1,2$, the counting of parameters to be tuned
is just {\em increased by one} relative to SU(3), due to the
additional quartic coupling $\lambda_7$ in \myref{su4lam} above.
For $N_f=2$ it has the form
\beq
\lambda_7 [ \e_{IJ} \e_{KL} \e^{ijk\ell} p_{i I} p_{j J} q_{k K} q_{\ell L} + \text{c.c.} ] .
\eeq
However, that coupling is forbidden for $N_f>2$ since
it will not be $SU(N_f)_p \times SU(N_f)_q$ invariant.
Thus for $N_f>3$ the Lagrangian and parameter counting
is identical to SU(3).

\subsection{$SU(N>4)$ gauge theory}
Here the form of the Lagrangian is just as in
SU(3).  The flavor specifications, depending on $N_f$
are likewise identical.

\subsection{Summary}
In summary, SQCD contains $\ord{10}$ fine-tunings
in each case.  For most of the theories, all of these
tunings are bosonic and can be done offline.  The exception
was SU(2) with $N_f=1$, where two Yukawa parameters must
be adjusted.  Setting aside that case, tuning between eight
and fourteen couplings on bosonic operators will pose a significant
challenge, even with the multicanonical reweighting
techniques that we discuss below.  A careful bootstrapping
method, from small to large lattices, will be necessary
in order to properly locate the critical parameter
values in such a large parameter space.  As mentioned
above, it is best to begin with weak couplings on
a small lattice, where lattice perturbation theory
should be a useful guide.  High statistics studies will
be required in order to constrain such a large number of
parameters.  Tuning against the \susy\ Ward identities,
as will be described for the $\Ncal=4$ SYM case in the
next section, also requires adjustment of mixing coefficients
for bare operators appearing in the supercurrent.
The problem appears daunting, and could only work
if an automated, recursive simulate/search strategy
is employed.

%%%%%%%%%%%% Lattice N=4 SYM ( N4GW ) %%%%%%%%%%%%%%%%%%

\mys{Lattice N=4 SYM}
\label{neq4}
In this section we describe a lattice formulation of four-dimensional
$\Ncal=4$ SYM\footnote{A review of the continuum theory,
its superconformal representations and the
AdS/CFT correspondence is given in \cite{D'Hoker:2002aw}.}
that may be within reach of practical simulations \cite{n4gw}, when combined with
the multicanonical methods described in \S\ref{mcrw}.  As for SQCD, we use GW fermions
to avoid gluino masses.  Just as important, the
GW fermions provide for an implementation of the 
global $SU(4)_R$ symmetry, which is chiral in how it couples
fermions and scalars.  The continuum chiral
$SU(4)_R$ symmetry is replaced by a lattice
generalization.  As will be seen, this symmetry
limits the number of counterterms that must be fine-tuned
in important ways.  As was
the case in SQCD, only bosonic operators require fine-tuning;
all tunings can be done ``offline'' by a
Ferrenberg-Swendsen \cite{Falcioni:1982cz,Ferrenberg:1988yz,Ferrenberg:1989ui} 
type reweighting, exploiting multicanonical
simulations to greatly broaden the parameter
space that can be scanned offline.  This aspect
of the theory will be described in detail
in \S\ref{mcrw}.

\subsection{Lattice Action}
\label{lata}
The continuum field content corresponds to $SU(N_c)$ Yang-Mills
coupled to scalars and fermions in an $SU(4)$ symmetric way.
Typically one writes the global symmetry as $SU(4)_R$,
where the $R$ denotes a symmetry that does not commute
with the generators of supersymmetry.
There are four massless Majorana fermions.
The left-handed components
transform in the fundamental ${\mathbf 4}$ representation
of $SU(4)_R$ and 6 real scalars in the ${\mathbf 6}$
representation (antisymmetric tensor).  If it were not for the Yukawa couplings,
we could formulate the theory instead in terms of two
Dirac fermions, which would simplify matters with respect to the GW
formulation.  However, the chiral 
$\psib_R \phi \psi_L \sim {\bf 4}_R \cdot {\bf 6} \cdot {\bf 4}_L$
Yukawa couplings require that we decompose the fermions into four
the left- and right-handed Majorana fermion components,
which are related to each other by charge conjugation.
(Note that $\psi_R \sim {\bf{\bar 4}}$ so that $\psib_R \sim {\bf 4}$.)

The six real scalars 
will be expressed with a single index $\phi_m$, $m{=}1{\dots}6$, or
composed into $SU(4)_R$
Weyl matrices:  $\phi_{ij}{=}\phi_m \sighat_{m,ij}$ and
$\phi^{ij} {=} \phi_m \sbarhat^{ij}_m$, where 
$\sighat$'s are just $SU(4)_R$ Clebsch-Gordon coefficients
involved in forming the singlet associated with ${\bf 4}_R \cdot {\bf 6} \cdot {\bf 4}_L$,
or equivalently ${\bf {\bar 4}} {\ni}\, {\bf 6} {\otimes} {\bf 4}$. 
They are most easily determined by dimensional reduction from ten
dimensions, or by recognizing them as the six-dimensional Weyl
matrices that are the building blocks of the six-dimensional Dirac gamma matrices.

The Euclidean continuum action is
\beq
S && = \frac{1}{g^2}\tr \bigg\{ \half G_{\mu \nu}G_{\mu \nu} +(D_\mu\phi_m)^2
+\bar{\psi}_i \nott{D}\psi_i \ddd
+\sqrt{2}\bar{\psi}_i\left(\phi^{ij} P_L{-}(\phi^{ij})^*P_R\right)\psi_j
+[\phi_m,\phi_n][\phi_m,\phi_n] \bigg\}.
\label{cane4}
\eeq
The $SU(4)_R$ preserving scalar lattice action must allow for generic
coefficients and non-\susyc\ terms, so that the \susy-restoring
counterterms can be tuned.
The quartic interaction terms in the SU(2) and SU(3) case are:
\beq
 \lambda_1 \tr \phi_m \phi_n \phi_m \phi_n
+ \lambda_2 \tr \phi_m \phi_m \phi_n \phi_n .
\label{stro}
\eeq
Comparing to \myref{cane4}, we see that classically
\susy\ corresponds to 
\beq
\lambda_1 = 1/g^2, \quad \lambda_2= -1/g^2.
\eeq
In the case of $SU(N_c>3)$, a total of four quartic terms
should be included, both the operators \myref{stro} as
well as
\beq
\lambda_3 \tr \phi_m \phi_n \tr \phi_m \phi_n +
\lambda_4  \tr \phi_m \phi_m \tr \phi_n \phi_n.
\label{strp}
\eeq
For SU(2) and SU(3) these can be eliminated in favor of the single trace
operators \myref{stro} using algebraic identities.  A scalar mass
term must also be included:
\beq
\half m^2 \tr \phi_m \phi_m.
\eeq
Regarding the kinetic term $(D_\mu\phi_m)^2$, one could use
a naive gauge covariant nearest neighbor approximation.  On the
other hand, it has been seen in many previous studies that taking
$D_\mu$ to be related to the fermion operator is advantageous
to reducing \susy-violating artifacts, presumably due to
degeneracies of modes in the UV where weak coupling 
applies \cite{Golterman:1988ta,Giedt:2004qs,Giedt:2004vb,Giedt:2005ae,Giedt:2005iq}.
Obviously such an implementation would be more demanding
numerically, since one uses the GW operator in the
scalar sector.  On the other hand, the advantages that
might come in reducing lattice artifacts may well make it
worth the effort.

The precise type of Ginsparg-Wilson~\cite{Ginsparg:1981bj} 
fermion to be used, 
be they domain wall~\cite{Kaplan:1992bt} or 
overlap~\cite{Neuberger:1997fp}, is not
important for the considerations here.
However, it has been argued that 
naive lattice Yukawa terms lead to inconsistencies in either the 
chiral or Majorana projections (depending on how the Yukawas are 
transcribed to the lattice)~\cite{Fujikawa2002}.
Following L\"uscher \cite{Luscher:1998pqa}, and Kikukawa and Suzuki
\cite{Kikukawa:2004dd}, we introduce auxiliary fermions
$\Psi$.  The fermionic lattice action is
\beq
S_{\text{F}} &=& \sum_x \tr\big\{ \bar\psi_i D \psi_i - \bar\Psi_i\Psi_i
\ddd \quad + y \sqrt{2} (\bar\psi{+}\bar\Psi)_i\left(\phi^{ij} P_L-(\phi^{ij})^*P_R\right)
(\psi{+}\Psi)_j\big\}.
\label{eqn:auxaction}
\eeq
where $D$ is the GW operator.
This action possesses an exact $SU(4)_R$ symmetry, with the scalars
transforming as in the continuum and the fermions transforming according
to
\beq
\hspace{-.2in}\begin{array}{l}
\delta\psi/i\eps=(T\hat{P}_{\!L}{-}T^*\hat{P}_{\!R})\psi \,, \\
\delta\Psi/i\eps=(T{+}T^*)\gamma_5 D \psi 
+ (TP_{\!L}{-}T^*P_{\!R})\Psi,\\ 
\delta\bar\psi/i\eps=\bar\psi\, (T^*P_{\!L}{-}TP_{\!R})
+ \bar\Psi (T{+}T^*)\gamma_5 \,, \\
\delta\bar\Psi/i\eps=-\bar\Psi\, (TP_{\!L}{-}T^*P_{\!R}) \,.
\end{array}
\label{latchi}
\eeq
Here $\hat{P}_{L/R}\equiv \half (1{\pm}\gfivehat)
= \half (1{\pm}\gfive(1{-}2 D ))$
are the lattice modified chiral projection operators, $T$ is the 
generator of $SU(4)_R$ in the fundamental ($\mathbf 4$)
and we have suppressed the $SU(4)_R$ indices.
Hence $(\psi{+}\Psi)$ and $(\bar\psi{+}\bar\Psi)$ transform like the
continuum $\psi,$ $\bar\psi$ fields. 

This auxiliary fermion method preserves the $R$-symmetry exactly and keeps
the Yukawa terms ultralocal. It is also consistent with the Majorana
decomposition, so the fermionic determinant is an exact square; taking
its square root to implement the Majorana nature of the fermions retains
locality.  The cost is the introduction of an extra fermionic excitation
$\Psi$, which is however nondynamical with $\ord{a^{-1}}$ mass, so it
decouples from the theory in the continuum limit.

In the case of $\phi=0$, it is known that the
overlap operator $D$ is non-negative; in particular,
$\det D \geq 0$.  If the domain wall fermion approximation
is used, then $\det D > 0$ for this case.  One
can ask what happens to this positivity feature
for $\phi \not= 0$.
It is easy to see that the fermion measure is real.
In the field space $(\psi,\Psi)$ the fermion matrix
has the $2 \times 2$ block form:
\beq
{\cal M} = \begin{pmatrix} D + M_Y & M_Y \cr
M_Y & M_Y-1 \end{pmatrix}, \quad
M_Y = y \sqtw \( \phi^{ij} P_L - (\phi^{ij})^* P_R \) .
\eeq
Since $\gamma_5 D^\dagger \gamma_5 = D$ and
similarly for $M_Y$, we have
\beq
(\det {\cal M})^* = \det {\cal M}^\dagger
= \det \gamma_5 {\cal M}^\dagger \gamma_5
= \det {\cal M} .
\eeq
The sign of the determinant may fluctuate.
In fact, to agree with some results from the
continuum (or really the zero-dimensional reduction---matrix models)
we know that it must \cite{Krauth:1998xh}.  Thus a sign problem
in the lattice theory may reflect continuum
dynamics.  One of the interesting
questions in a lattice study is the extent
to which this correlates with motion through
the nontrivial moduli space in $\Ncal=4$ SYM.
In a simulation, the sign fluctuations will have to be accounted
for by monitoring the low-lying eigenvalues of the fermion matrix,
which can be computed efficiently.

\subsection{Tuning to the supersymmetric theory}
\label{stng}
Our goal is to nonperturbatively tune the lattice action 
such that the IR description is a good approximation
to \NfourSYM, with errors that are $\ord{a}$ and the
lattice spacing $a$ much
smaller than the scales of interest. 
Due to operator mixing there is a nontrivial matching between 
the lattice and effective IR theories.  All
relevant and marginal terms consistent with lattice symmetries
will appear in the infrared, except at special points in
bare parameter space.  We can arrive at the desired special
point, $\Ncal=4$ SYM, by introducing the \susy-violating operators
into the bare action and fine-tuning counterterms. 
These counterterms fall into three categories:
a scalar mass term, a Yukawa term, and two or four scalar quartic terms,
depending on the number of colors for the gauge group,
restricted here to $SU(N_c)$.
As has already been mentioned, if $N_c \leq 3$ then only two quartic terms need
to be included, Eq.~\myref{stro}, while
for $N_c > 3$ two more quartic terms must be introduced, Eq.~\myref{strp}.
As in the SQCD discussion above, rescaling the Yukawa term
can be accomplished through a rescaling of the scalar kinetic term.
Therefore in the lattice theory the scalar kinetic term
should be taken to have a general coefficient $Z_\phi$ that
is to be tuned nonperturbatively.  We will describe the
multicanonical reweighting method of fine-tuning the $\Ncal=4$
SYM lattice theory in more detail in \S\ref{mcrw} below.
This will include a discussion of mixing coefficients
that must be measured in the supercurrent in order to
use \susy\ Ward identities in the fine-tuning process.

We now comment  on the significance of preserving
the chiral symmetries of the theory, albeit in the
lattice-modified form \myref{latchi}.  
For this purpose suppose we were to formulate
the fermionic part of the theory instead as 
\beq
S_{\text{F}} &=& \sum_x \tr\big\{ \bar\psi_i D_w \psi_i 
+ \frac{y}{\sqrt{2}} \bar\psi_i \left( \phi^{ij} P_L - (\phi^{ij})^* 
P_R \right) \psi_j \big\}
\label{wfe}
\eeq
with $D_w$ the Wilson-Dirac operator.  Then due to the
explicit violation of chiral symmetry, a mass correction
\beq
m \psib_i \psi_i = m \( \psib_{Li} \psi_{Ri} + \psib_{Ri} \psi_{Li} \)
\label{so4m}
\eeq
would be generated.  Note that the Majorana condition $\psi_i^c = \psi_i$
implies that $\psi_{Li}$ and $\psi_{Ri}$ are in conjugate representations:
\beq
P_R \psi^c = C \psib_L^T = \psi_R, \quad
P_L \psi^c = C \psib_R^T = \psi_L,
\eeq
Comparing to \myref{so4m}, we see that
the mass term consists of $4 \cdot 4$ and
$\bar 4 \cdot \bar 4$ couplings, 
so that only the real subgroup $SO(4)$ of the
original $SU(4)_R$ symmetry is preserved in
this Wilson-Dirac formulation.  Futhermore,
this mass term converts $\psi_{Li}$ into
$\psi_{Ri}$ (thus mixing the $4$ and $\bar 4$
representations of the $SU(4)_R$), so
we will radiatively generate new Yukawa couplings
\beq
y' \sqrt{2} \psib_i \left( \phi^{ij} P_R - (\phi^{ij})^* P_L \right) \psi_j ,
\eeq
where chiralities have been swapped relative to
the \susyc\ Yukawa term in \myref{wfe}.

On one hand, the $SO(4)$ that is preserved does limit
the number of parameters that must be fine-tuned.  We have
just two more, $m$ and $y'$, than in the GW case.  On the
other hand, these are additional fermionic counterterms,
and we cannot use the trick of rescaling the scalar
kinetic term, since that freedom has already been
exploited for tuning the parameter $y$.  So, we face the
problem that two new terms of a very problematic
type are present because we have not preserved the
chiral $SU(4)_R$, but only a real subgroup.  This will
also pose a problem for tuning of Ward identities, because
more operators can mix with the supercurrent, due
to the reduced symmetry of the lattice theory.  That
translates into additional mixing coefficients that
must be measured nonperturbatively.

\subsection{Discussion}
We have used GW fermions with chiral $SU(4)_R$ invariant Yukawa couplings 
following the method of \cite{Luscher:1998pqa,Kikukawa:2004dd}
to reduce the counterterms in an important way.
Because the counterterm tuning bosonic, it can be done offline.
In \S\ref{mcrw} we will explain how to alleviate the {\it ensemble 
overlap problem} by taking a multicanonical approach, 
flattening the distribution with respect to the
parameters that are to be scanned.

We invite the reader to contemplate the following
circumstance, which is interesting since it is
radically different from what occurs in lattice QCD:
since \NfourSYM\ is conformal, the 
continuum limit is not a weak coupling limit $g \to 0$. 
What is known about such lattice field theories?
Perhaps the best starting point is the class of
two-dimensional models with an IR fixed point
that have been extensively
on the lattice.  
But $\Ncal=4$ SYM should have not
merely an \IRFP\ (as is supposed to occur in some 4d gauge
theories that have been studied recently on the
lattice \cite{Appelquist:2007hu,Shamir:2008pb,Appelquist:2009ty}), but a whole critical
line of fixed points.  The well-known two-dimensional analogy
is the XY model (planar spin model, O(2) model, etc.). 
In that case,  each $\kappa \geq \kappa_c$  leads to
a conformal field theory (CFT) in the IR, but it is in fact a continuous family
of CFTs, with scaling exponents (anomalous dimensions) that depend on 
$\kappa$.\footnote{Recall that large $\kappa$
corresponds to low temperature, which is precisely
where order is anticipated.  Of course in the XY
model it is just {\it algebraic order,} due
to the absence of spontaneous
symmetry breaking in two dimensions.}  
The same should be true for $\Ncal=4$ SYM:  each value of
the continuum gauge coupling $g_{cont.}$ corresponds to a CFT,
but the anomalous dimensions of {\it non-chiral,} 
or {\it non-BPS,} quantities will
depend on the value of $g_{cont.}$. 
The lattice formulation should 
also have that feature, though the relation between
the lattice coupling $g_{latt.}$ and the continuum one $g_{cont.}$
must be established through detailed calculations.

Our proposal should certainly work
at very weak coupling $g_{latt.} \lappeq g_{cont.} \ll 1$, 
where one knows that the IR description will be
in terms of the same degrees of freedom as one puts on the lattice.
Here it is important that, due to the fine-tuning,
one starts with $g_{latt.}$ at the lattice scale and
the theory flows into the \IRFP\ such that
the coupling remains weak, terminating with the value $g_{cont.} \ll 1$.
Of course in such a situation perturbation theory is reliable
and there is no need to use lattice discretization.  Nevertheless,
this is a useful reference point for the lattice parameters
that must be fine-tuned.

But is it guaranteed that one can find lattice parameters which
correspond to the strongly coupled continuum theory?
No definitive answer to this question will be offered here,
though we certainly have our own opinion.  We only
see room for four possibilities:
\ben
\item No lattice field theory
can describe strong $\Ncal=4$ SYM, but there is another,
more sophisticated nonperturbative formulation that can do the job.
\item Strong $\Ncal=4$ SYM does not really exist---it
is just a continuum field theorists' fantasy.
\item A lattice field theory exists that will do the job of describing
the strongly coupled theory, but it is not 
the one formulated  here.
\item Along a line of points in the parameter
space of the lattice theory described here,
strongly coupled $\Ncal=4$ SYM emerges in the IR.
\een

Our opinion is that first two possibilities are
virtually impossible.  In particular, there is convincing
evidence from the AdS/CFT correspondence that
strong $\Ncal=4$ SYM corresponds to a weakly
coupled supergravity theory, and that it is
perfectly consistent \cite{Maldacena:1997re,
Gubser:1998bc,Witten:1998qj}.   In the gauge theory,
the anomalous dimensions of {\it BPS operators}  are
determined by the superconformal algebra and are protected
from renormalization.  Thus they can be computed at
arbitrarily weak coupling and then continued into
the strong coupling regime.  Thus certain operators
written in terms of elementary fields have a well understood
behavior in the IR.  Given that this is true,
it is hard to imagine why the elementary fields would
not provide ``the correct degrees of freedom'' in
terms of which to define the theory at strong coupling.

Furthermore, the renormalization
group perspective indicates that some lattice theory
should exist whereby the lattice artifacts can be
cancelled by irrelevant operators, leading to a {\it perfect}
lattice action.\footnote{Of course
this assumes that continuum power-counting
can be applied.  However, it is reasonable
to suppose that with sufficient effort one's
intuition in this regard can be proven, as
has been done recently for staggered fermions \cite{Giedt:2006ib}.}
Such a perfect lattice action theory
would only differ from the one we are proposing
by (naively) irrelevant terms.  The fine-tuning approach 
that we are advocating could only fail (Option 3 in the list above)
if the naively irrelevant operators that appear in
the perfect lattice action theory turn out to
be relevant or marginal in the IR as one moves to stronger
lattice gauge couplings.  

However, we believe that Item 4 is
the true state of affairs, though at this point it
is a matter of speculation.
The lattice will introduce artifacts that break conformal symmetry
by irrelevant operators.  This will be characterized by the lattice
spacing $a$.  Thus starting at the lattice scale, there is
a renormalization group flow into the IR fixed point
where the theory becomes conformal.  Equivalently, one must
look at distances $x \gg a$ in order to see the conformal
field theory behavior.

For instance, correlation functions over a distance $x$
will depend on $a$ and the linear size of the lattice $L$, where
the latter serves as an IR regulator.  They will have the general
form
\beq
C(x) = |x|^{-2h}\( 1 + \ord{a/x,x/L} \)
\label{corrwe}
\eeq
where $h$ is the scaling dimension of the operator and the ${\cal O} (a/x,x/L)$
correction carries the scaling violation due to irrelevant lattice
artifacts, as well as the finite size
corrections.  One cannot simply take $L \to \infty$ to remove these,
since the theory needs an IR regulator if it is to make
sense, under the assumption that it flows into an IR fixed
point.

Thus, suppose the properly tuned lattice action flows 
into an \IRFP.  There is
some scale $\ell_{\text{CFT}}$ beyond which further flow is negligible.
For $x > \ell_{\text{CFT}}$, the behavior is that of a CFT.
In order to capture this regime, 
it is important that $\ell_{\text{CFT}} \ll L$.
We know that we have a fine lattice (small lattice
spacing $a$) if 
\beq
\ell_{\text{CFT}} / a \gg 1 .
\label{lcfta}
\eeq
To identify the distance scale at which conformality
sets in, one could monitor the running gauge coupling
using Schr\"odinger functional and step-scaling
methods \cite{Luscher:1992an,Luscher:1993gh}, as has been
done in \cite{Appelquist:2007hu,Shamir:2008pb,Svetitsky:2008bw,Appelquist:2009ty}.

To have a strong CFT, one should look for
$\ord{1}$ anomalous dimensions through measurements
of critical exponents.  Admittedly this would be
an enormous challenge to accomplish
through the standard method of finite-size scaling studies.
This is because the theory is four-dimensional with GW
fermions and fine-tuning that grows exponentially
more difficult as the volume is increased.  On the other
hand, one can fantasize that it may be possible to 
extract information from small volume studies, 
analogous to the $\e$ regime methods that are used in QCD.  If one
were to accomplish this super-human task, find $\ord{1}$
anomalous dimensions, and verify that all of the $SU(2,2|4) \times SU(4)_R$
Ward identities are satisfied, then it is hard to see
how one has anything other than strong $\Ncal=4$ SYM,
based on universality arguments.
The capstone of such a study would be to match the
anomalous dimensions to those predicted by the AdS/CFT correspondence.

In general, lattice theories have a phase structure that is
richer than the continuum theory that they are intended to define.
A well-known example is SU(2) gauge theory with a mixed
fundamental/adjoint Wilson action \cite{Greensite:1981hw,Bhanot:1981eb}.
Phase boundaries exist
and some regions of the lattice parameter space have only
$\ord{a}$ correlations.  These
are ``lattice phases'' separated from the phase with a continuum
limit by a {\it bulk transition.}  
This will happen in the $\Ncal=4$ SYM SU(2) lattice theory, since several
adjoint fermions are present; under coarse-graining they will generate
adjoint Wilson action terms:
\beq
S_{eff} \ni -\beta_A \sum_x \tr U_{\mu \nu}^{( \text{adj} )}(x) .
\eeq
Indeed, this has recently 
been observed to give rise to a bulk transition in SU(2) lattice gauge
theory with four adjoint Majorana fermions \cite{Catterall:2008qk}.
Similar findings have been reported for SU(3) with sextet fermions
\cite{DeGrand:2008kx,DeGrand:2008dh}.  
The situation in $\Ncal=4$ SYM lattice theory with gauge group SU(2) will be similar to
what was found in \cite{Catterall:2008qk}:  a continuum
phase exists only if the bare lattice coupling $\beta=4/g^2$ is sufficiently weak
(e.g., $g^2 \lappeq 2$ in \cite{Catterall:2008qk}).  

The point here is that even though
the target theory is one in which the coupling
constant does not run, in the lattice theory irrelevant
operators cause important renormalizations at
short distances so that the true strength of the IR coupling 
must be determined by a detailed study of the long distance 
physics.

\mys{Fine-tuning with multicanonical reweighting}
\label{mcrw}
Multicanonical methods \cite{Binder,Baumann:1986iq,Berg:1991cf}
combined with ``Ferrenberg-Swendsen 
reweighting'' \cite{Falcioni:1982cz,Ferrenberg:1988yz,Ferrenberg:1989ui}
[refered to here as multicanonical reweighting (MCRW)] have proven
to be a powerful tool for maximizing the usefulness
of Monte Carlo simulations over a range of parameter space
much wider than was actually simulated. 
For instance, MCRW was applied in a study comparing $SU(2)$ and $SO(3)=SU(2)/Z_2$
lattice gauge theories \cite{deForcrand:2002vx,deForcrand:2002vs}.  It
was found to dramatically flatten the distributions
with respect to three parameters, twists on gauge fields
at the spatial boundaries.  Another successful
application of MCRW 
consists of lattice results for the electroweak
phase transition \cite{kajantie,moore_electroweak_2000}.

We will begin by describing  
MCRW generally, followed by a presentation
of how it would be applied to the $\Ncal=4$ lattice SYM
that was described in the previous section.

\subsection{Preliminaries}
Suppose we perform a Monte Carlo simulation at one value $m_0$ of the scalar mass $m$,
so that the configurations sample the distribution determined
by the action 
\beq
S(m = m_0) = S(m=0) + \half \int d^4 x ~ m_0^2 \phi^2(x) .
\label{m0act}
\eeq
Following the ``Ferrenberg-Swendsen reweighting'' method 
\cite{Falcioni:1982cz,Ferrenberg:1988yz,Ferrenberg:1989ui}
one can use the following
{\em reweighting identity} to compute the expectation value $\vev{ \Ocal }_m $ of an operator
$\Ocal$ for the distribution with a mass $m$:
\beq
\vev{ \Ocal }_m = \frac{ \vev{\Ocal e^{- \Delta S(m)} }_{m_0} }{ \vev{ e^{- \Delta S(m)} }_{m_0} }
\approx \frac{\sum_{C \in F(n)} 
\Ocal_C  \exp \[ -\half ({m}^2-m_0^2) \int d^4 x ~ \phi^2_C \] }
{\sum_{C \in F(n)} \exp \[ -\half ({m}^2-m_0^2) \int d^4 x ~ \phi^2_C \] } .
\label{ilte}
\eeq
In the first equality $\vev{ \cdots }_{m_0}$ is the expectation value with
respect to the canonical distribution corresponding to \myref{m0act} and
\beq
\Delta S(m) = \half ( m^2-m_0^2) \int d^4  x ~ \phi^2
\eeq
is the shift in the action when the mass is changed.  In the second,
$\int d^4x ~ \phi^2_C$ and $\Ocal_C$ are the mass term and
operator evaluated on configuration $C$
and $\sum_{C \in F(n)}$ is the sum over the distribution
$F(n)$ of $n$ configurations generated in the Monte
Carlo simulation.  These of course provide a finite
ensemble that approximates the canonical distribution
corresponding to \myref{m0act}.  The advantage of this
approach is that one need only perform a single simulation
at mass $m_0$, storing the values of $\int d^4x ~ \phi^2_C$ and 
$\Ocal_C$ for each $C$, and then $\vev{ \Ocal }_m$ can
be computed for a swath of the parameter space $m$ without
having to perform any new simulations.  Typically the
time for this ``offline'' calculation is negligible compared
to that of the simulation.

Unfortunately, the regime of utility for this technique is limited
by the {\it overlap problem,} in a way that often worsens
exponentially in the spacetime volume.  For instance,
suppose the theory \myref{m0act} has a quartic interaction
and a critical mass-squared $m_c^2$ such that for $m^2 < m_c^2$ there
is spontaneous symmetry breaking.  If we simulate with
$m_0^2 > m_c^2$ then the field is exponentially weighted
toward $\int d^4x ~ \phi^2 \approx 0$.  Now suppose we
attempt to reweight to $m^2 < m_c^2$.  In that case $-(m^2-m_0^2)>0$
so that the exponential weight factor in \myref{ilte} is
minimal at $\int d^4x ~ \phi^2 \approx 0$.  The ensemble
that is generated in the Monte Carlo simulation will have
exponentially few configuration in the regime where
$\int d^4x ~ \phi^2$ is far from zero and $e^{-\Delta S(m)}$
is large.  Because we will have very few representatives
of configurations with the largest weight $e^{-\Delta S(m)}$,
and most members of the ensemble have very small weight, fluctuations
will be large and huge samples are 
required in order to have acceptable errors.  
The mismatch of the distributions gets worse as 
the number of lattice sites increases, because
the exponent is extensive (i.e., scales like the spacetime
volume $L^4$).

\begin{figure}
\begin{center}
\includegraphics[width=5in,height=3in]{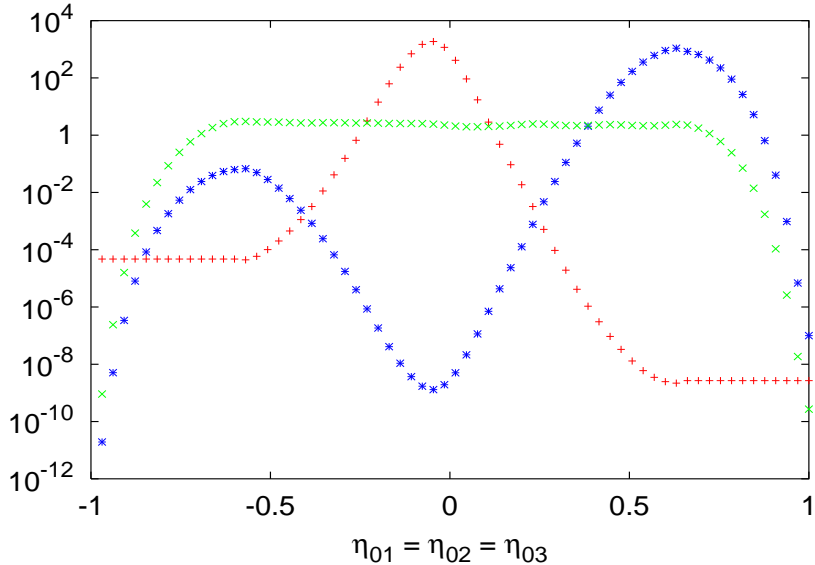}
\caption{Fig.~4 of \cite{deForcrand:2002vs}, with
permission.  This shows the unweighted distribution (blue burst),
multicanonical reweighting function (red plus) and the
reweighted distribution (green x). \label{defor}}
\end{center}
\end{figure}

As a concrete example, we reproduce in Fig.~\ref{defor}
a figure from \cite{deForcrand:2002vs}.  It shows that in
the range of a three-dimensional parameter space the
ordinary canonical Monte Carlo distribution varies by 15
orders of magnitude.  This is for an $8^3 \times 4$ lattice,
which is still relatively small.

In a number of contexts 
the technique of {\it multicanonical reweighting} \cite{Binder,
Baumann:1986iq,Berg:1991cf}
has been found to ameliorate the overlap problem.
One replaces $S$ with  
\beq
S_{MCRW}=S+W[\Ocal_1, \Ocal_2, \ldots], 
\label{srww}
\eeq
where $W[\Ocal_1, \Ocal_2, \ldots]$ is a
carefully engineered function of some small set of observables.
For instance in the $\Ncal=4$
SYM case $W$ will be a function of $\int d^4 x ~ \phi^2$, the distinct 
quartic terms $\int d^4 x ~ \phi^4$ and the kinetic term 
$\int d^4 x ~ (D \phi)^2$.
The (reweighted) expectation value of an observable in the distribution corresponding
to $S_{MCRW}$ is:
\beq
\vev{\cal O} = \frac{\sum_{C \in F(n)} \Ocal_C \; e^{ W[\Ocal^C_1,...] }}
{\sum_{C \in F(n)} e^{ W[\Ocal^C_1,...]} } .
\label{rwww}
\eeq

Since the $e^W$ factor in \myref{rwww} just cancels the $e^{-W}$ Boltzmann factor
coming from \myref{srww}, one might wonder why it is introduced in the first place.
The point is that the additional Boltzmann factor $e^{-W}$ in effect
produces a weighted average over a continuum of canonical ensembles 
(hence the appelation ``multicanonical'') such that there is
a good overlap with the distribution that one is
reweighting to.  The challenge is to design a $W$ such that
sampling is flattened over the range of observables one is interested in.

We return to Fig.~\ref{defor}, taken from \cite{deForcrand:2002vs}.
It shows that
the multicanonical Monte Carlo sampling distribution is flat in
the range of three-dimensional parameter space between the
peaks, where the
ordinary canonical Monte Carlo distribution varies by twelve
orders of magnitude.  The reweighting function $W$ was
represented by a numerical table, composed of the inverse
density of states with respect to the tuned parameters.
This is for an $8^3 \times 4$ lattice,
which is still relatively small, and it indicates
that $\ord{10^{12}}$ more samples would be required
in the canonical Monte Carlo approach in order
to scan a comparable range of parameter space
by ordinary Ferrenberg-Swendsen reweighting techniques.  Working on lattices
of size, say, $16^4$, would make the overlap problems of
the canonical distribution many orders of magnitude worse.  But lattices
of this size and larger are needed in order to extract continuum
behavior from the lattice.  On the other hand, it is
not known how difficult the overlap problem is in the
two types of \susyc\ models considered above, SQCD and
$\Ncal=4$ SYM.

As another example, in studying first order phase transitions 
(e.g., \cite{kajantie}), one chooses
$\Ocal_1$ to be the order parameter of the transition; in a model with a
scalar field, typically $\Ocal_1 = \int d^4 x \phi^2$.  One tunes $W[\Ocal_1]$ to
cancel the nonperturbative effective potential for this operator, so that
the Monte Carlo simulation samples evenly in $\Ocal_1$.  This enhances
statistics for configurations intermediate between the phases.
In the mass scan example of Eq.~\myref{ilte}, one has
\beq
\vev{\Ocal} = \frac{\sum_{C \in F(n)} \Ocal_C \exp \( W[\int d^4 x ~ \phi_C^2] 
- \half (m_2^2-m_1^2) \int d^4 x ~ \phi^2_C \) }
{\sum_{C \in F(n)} 
\exp \( W[\int d^4 x ~ \phi_C^2] -\half (m_2^2-m_1^2) \int d^4 x ~ \phi^2_C \) } .
\label{omex}
\eeq
In this way, wherever the exponential in \myref{omex} 
happens to be at its maximum,
a large number of configurations will be generated, due to
the flat distribution with respect to $\int d^4 x ~ \phi^2$.

Two approaches exist for engineering a good function $W$.
\bit
\item[(1)]
One can employ a bootstrap method that iterates between Monte Carlo 
simulation and adjustments to $W$.  For
instance a numerical tabulation of density of
states $\rho$ may be obtained from a canonical simulation,
as was done in \cite{deForcrand:2002vx,deForcrand:2002vs}.  Schematically,
one obtains a histogram estimate
of $\rho(\Ocal_1)$ for an operator value range $\Ocal_1$ range $\Ocal_{1,\text{min}}
\leq \Ocal_1 \leq \Ocal_{1,\text{max}}$.  This provides
an initial version of $W$, through $W \equiv 1/\rho(\Ocal_1)$.
If necessary, the process can be repeated to refine the table.
\item[(2)]
Iterative or stochastic searches may be used to
optimize $W$ with respect to a predetermined parameterization in a small volume.
Performing this at two different small volumes then provides
an extrapolation estimate for $W$ in the next largest volume,
which can then be refined through another search.
\eit

\subsection{Application of MCRW to $\Ncal=4$ SYM}
Numerical studies of MCRW for $\Ncal=4$ SYM have yet
to be attempted, though the groundwork for this
effort was laid in \cite{n4gw}.  Here we will review
those findings, as an illustration of the MCRW approach
to lattice \susy.

For $N_c < 4$, the reweighting function $W$ will depend on the four bosonic
contributions to the action
\beq
&&\Ocal_1 = \int d^4 x ~ \tr D_\mu \phi_m D_\mu \phi_m,
\quad
\Ocal_2 = \int d^4 x ~ \tr \phi_m \phi_m,
\ddd
\Ocal_3 = \int d^4 x ~ \tr \phi_m \phi_n \phi_m \phi_n,
\quad
\Ocal_4 = \int d^4 x ~ \tr \phi_m \phi_m \phi_n \phi_n.
\label{ocdf1}
\eeq
For $N_c \geq 4$ we must also include the double-trace operators
\beq
\Ocal_5 = \int d^4 x ~ \tr \phi_m \phi_n \tr \phi_m \phi_n,
\quad
\Ocal_6 = \int d^4 x ~ \tr \phi_m \phi_m \tr \phi_n \phi_n.
\label{ocdf2}
\eeq

A finite sampling of the multicanonical ensemble described by
\beq
Z_{MC}=
\int [dA ~ d\psi ~ d\phi] \exp \[ - S(Z_\phi^{(0)},m_0^2,\lambda_i^{(0)}) 
- W( \{ \Ocal_i \} ) \]
\label{zmcdi}
\eeq
is performed by Monte Carlo simulation,
where $Z_\phi^{(0)},m_0^2,\lambda_i^{(0)}$
are the coefficients of the operators
\myref{ocdf1} and \myref{ocdf2} that appear
in the action.  In the simulation the RHMC  algorithm
\cite{rhmc98,rhmc03,rhmc04} should be used as this is currently
the best available approach for dynamical fermions.
Of course one replaces the continuum gauge fields $A_\mu(x)$ with
lattice link fields $U_\mu(x)$, and the fermions are
replaced with pseudofermions $\chi^i(x)$, with a
corresponding reformulation of the lattice action,
in the usual way.
One then computes reweighted expectation  values
of quantities $\Ocal$ for
a different set of parameters $Z_\phi, m^2, \lambda_i$
{\it via} the relation
\beq
&&\vev{\Ocal}_{MCRW} = \frac{ \vev{\Ocal \exp \[ -\Delta S + W( \{ \Ocal_i \} ) \]}_{MC} }
{\vev{\exp \[ -\Delta S + W( \{ \Ocal_i \} ) \] }_{MC} }, \ddd
\Delta S( Z_\phi,m^2,\lambda_i | \{ \Ocal_i \} )  
\equiv S ( Z_\phi,m^2,\lambda_i | A, \psi, \phi )
- S ( Z_\phi^{(0)},m_0^2,\lambda_i^{(0)} |  A, \psi, \phi )  ~~~
\label{mcrwe}
\eeq
where MC indicates the multicanonical expectation value
following from simulation with \myref{zmcdi}.
We have made it explicit that the change in action
$\Delta S$ depends only on the operators $\Ocal_i$
that were given in \myref{ocdf1} and \myref{ocdf2}.

The finite sample $F(n)$ generated from the Monte Carlo
simulation with distribution described by \myref{zmcdi} consists of a set of
configurations $C_1,\ldots,C_n$.  Thus for a given
$C$ the lattice fields $U_\mu(x),\chi^i(x),\phi_m(x)$
take values $U_\mu(C|x),\chi^i(C|x),\phi_m(C|x)$.
From this point of view the MCRW evaluation of
expectations values \myref{mcrwe} can be interpreted
as computations with partition function
\beq
Z_{MCRW} = \lim_{n\to\infty} \sum_{C \in F(n)} \exp \[ -\Delta S_C(\{ \Ocal_i \})
+ W_C( \{ \Ocal_i \} ) \]
\label{zmcrw1}
\eeq
where the subscripts $C$ inside the exponential
indicate that all fields are to be evaluated on
configuration $C$.

Next we define a density of states $n(\Ocal_1,\ldots,\Ocal_4) \equiv
n( \{ \Ocal_i \} )$
in the multicanonical distribution, where we have
specialized to the $N_c < 4$ case for notational
simplicity.  Let $\mathfrak{F} (\{ \Ocal_i \} )$
be any function of the operators.  Then:
\beq
&& \int [dA ~ d\psi ~ d\phi]~  e^{-S[A,\psi,\phi]-W[\{ \Ocal_i[\phi] \}]} ~
\mathfrak{F} (\{ \Ocal_i[\phi] \}) =
\ddd \int [dA ~ d\psi ~ d\phi] ~ e^{-S[A,\psi,\phi]
-W[\{ \Ocal_i[\phi] \}]} ~ \mathfrak{F} (\{ \Ocal_i[\phi] \})
\( \int \prod_i d\Ocal_i^{(0)} \delta( \Ocal_i[\phi]-\Ocal_i^{(0)} ) \)
\ddd \equiv
\( \int \prod_i d\Ocal_i^{(0)} \) n( \{ \Ocal_i^{(0)} \} )
\mathfrak{F} (\{ \Ocal_i^{(0)}[\phi] \})
\eeq
Thus we can write the reweighted multicanonical partition function
\myref{zmcrw1} as:
\beq
&& Z_{MCRW} = \( \int \prod_i d\Ocal_i \) ~ n( \{ \Ocal_i \} ) ~
w( Z_\phi,m^2,\lambda_i | \{ \Ocal_i \} )  
\ddd
w( Z_\phi,m^2,\lambda_i | \{ \Ocal_i \} )  \equiv
\exp \[ - \Delta S( Z_\phi,m^2,\lambda_i | \{ \Ocal_i \} ) + W( \{ \Ocal_i \}) \]  
\eeq

The engineering of $W( \{ \Ocal_i \} )$ has as its goal
the generation of ensembles $F(n)$ such that there is a
reasonable number of configurations $n( \{ \Ocal_i^{(0)} \} )$ with
large weight for a broad
patch in the parameter space $Z_\phi, m^2, \lambda_i$ that
we intend to scan in the fine-tuning process.
For a choice of $Z_\phi, m^2, \lambda_i$ within that patch,
what we therefore want is $n( \{ \Ocal_i^{(0)} \} )$ not
too small wherever $w( Z_\phi,m^2,\lambda_i | \{ \Ocal_i \} )$
has most of its support.
Of course one must also decide where the patches of
interest lie.  This  is
 best achieved through  the bootstrap method
described above (i.e., starting from small lattices
and weak couplings, where the counterterms can
be determined reliably using analytic methods).

\subsubsection{Effective potential}
\label{sstff}
\paragraph{Tuning the scalar mass term.}
We begin by considering the case where we only shift $m_0^2 \to m^2$
relative to the reference point $(Z_\phi^{(0)},m_0^2,\lambda_i^{(0)})$
of the multicanonical ensemble.  One sees from \myref{mcrwe}
that $\Delta S = (m^2-m_0^2) \Ocal_2$, where
$\Ocal_2$ is the mass
operator defined in \myref{ocdf1}.  
The gauge invariant effective potential in finite
volume is defined as follows:
\beq
e^{ -\Omega V_{\text{eff}}(m^2|A^2) } = \bigvev{ e^{ W( \{ \Ocal_i\} ) 
- (m^2-m_0^2) \Ocal_2 } ~ \delta \( A^2 - \frac{\Ocal_2}{\Omega} \) }
\label{vefeq}
\eeq
where $\Omega$ is the spacetime volume.  
Thus $A^2$ represents the mean value of
the squared scalar field $\tr \phi_m \phi_m$.

Now suppose we vary
\beq
m^2 \to m^2 + \Delta m^2.
\label{msqsh}
\eeq
In \myref{vefeq} we can use the $\delta$-function
as follows:
\beq
e^{- \Delta m^2 \Ocal_2 } \delta \( A^2 - \frac{\Ocal_2}{\Omega} \)
= e^{- \Delta m^2 \Omega A^2} \delta \( A^2 - \frac{\Ocal_2}{\Omega} \) .
\eeq
Since we can take $e^{- \Delta m^2 \Omega A^2}$ outside
the expectation value, it is clear that the
shift \myref{msqsh} changes $V_{\text{eff}}(m^2|A^2)$ by adding a
linear component:
\beq
V_{\text{eff}}(m^2+\Delta m^2|A^2) = V_{\text{eff}}(m^2|A^2) + \Delta m^2 A^2.
\eeq
Therefore, measuring $V_{\text{eff}}(m^2|A^2)$ immediately
determines how $\vev{ \tr \phi_m \phi_m}$ and 
$F_{MCRW} = -\ln Z_{MCRW}$ vary as a function
of $m^2$.

For example, suppose we measure $V_{\text{eff}}$
with $m^2=0$, and obtain a parameterization
\beq
V_{\text{eff}}(m^2=0|A^2) = r A^2 + u A^4 + \cdots
\eeq
where $\cdots$ represents terms higher order in $A^2$.
Then it is clear that the critical mass is $m^2 = -r$.

As an alternative, one can also locate the critical $m^2$ by looking for
the peak in the $\phi_m$ susceptibility
\beq
\chi(\phi) = \int d^4 x ~ \bigvev{  \tr \phi_m(x) \phi_m(0) }
\label{scsu}
\eeq
On the lattice, gauge-fixing will be necessary since
the fields are located at different sites $x$.
As with the determination of the effective potential,
this can be done offline using the reweighting techniques.
The peak of the susceptibility $\chi(\phi)$ with
respect to $m^2$ should agree with the point located by
the effective potential analysis that we
discussed in relation to \myref{vefeq}.

\paragraph{Tuning the quartic terms.}
Finding the quartic parameters that lead to additional
second-order behavior should be possible.  Indeed, 
it has been successfully achieved in
the context of the electroweak phase transition \cite{kajantie}.
In the $\Ncal=4$ theory one could for instance
look for peaks in the following susceptibility tensor:
\beq
\chi_{mn;pq}  &=&
\int d^4 x ~ \bigvev{ \Ocal_{m n}(x) \Ocal_{m n}(0) }_{\text{conn.}} ,
\nnn
\Ocal_{mn} &=& \tr \phi_m \phi_n - \frac{1}{6} \delta_{mn} \sum_k \tr \phi_k \phi_k,
\label{qsucp}
\eeq
where ``conn.'' denotes the connected correlation function.
On the basis of SO(6) and cyclic trace symmetries, one has
for the susceptibility:
\beq
\chi_{mn;pq} = c_1 ( \delta_{mp} \delta_{nq} + \delta_{mq} \delta_{np} )
+ c_2 \delta_{mn} \delta_{pq} .
\label{genqu}
\eeq
In the \susyc\ theory, the operator in \myref{qsucp} is the
chiral primary operator of the {\it supergravity multiplet,}
where the terminology arises from the AdS/CFT correspondence.
It is 1/2 BPS so that its conformal dimension $\Delta=2$ is protected,
and the susceptibility must reflect the fact that it transforms
as a $\mathbf{20'}$ of SO(6).  Taking these properties into
account, we have at the \susyc\ point the prediction
\beq
\chi_{mn;pq} \sim \( \delta_{mp} \delta_{nq} + \delta_{mq} \delta_{np} \) \ln (L/a),
\label{susqu}
\eeq
where $L$ is the linear size of the lattice, which serves
as an IR cutoff.  The term $c_2$ in \myref{genqu} vanishes
in the \susyc\ limit since $\Ocal_{mn}$ creates an
exact chiral primary state.  Of course we work here
with bare lattice operators and mixing will occur.
However, the logarithmically divergent susceptibility
should provide a clear signal of the additional
second order behavior associated with tuning to the \susyc\
point.  It is interesting to contrast this with the
scalar susceptibility of the {\it Konishi multiplet,}
$\Ocal_K = \sum_k \tr \phi_k \phi_k$, which according
to perturbation theory \cite{Anselmi:1996dd} has $\Delta_K > 2$,
and hence finite susceptibility with respect to $L \to \infty$
at fixed lattice spacing $a$.  Because it is non-BPS,
the AdS/CFT correspondence cannot be used to check the
conformal dimension $\Delta_K$ at strong coupling.  This
is one feature that might be probed with the lattice,
by examining the finite-size scaling of the corresponding
susceptibility.

Thus suppose that as we tune the quartic parameters $\lambda_i$ toward
the \susyc\ point, flat directions of the scalar potential
open up, revealing the moduli space of $\Ncal=4$ through
divergent scalar susceptibilities.  This presents numerical difficulties
and one would rather see the divergence as a limiting
behavior.  While it is true that lattice artifacts
regulate this divergence, it is nevertheless desireable
to have an independent knob that controls it.  Furthermore,
near the \susyc\ point the quartic potential can turn over, leading to a
runaway instability.  To regulate the runaway directions
and render susceptibilities finite, we propose to add
a sextic term
\beq
\mu_6 a^2 \int d^4 x ~ (\tr \phi_m \phi_m)^3
\label{irel6}
\eeq
to the potential.  Here, $a$ is the lattice
spacing, or in a continuum description, the inverse of the UV momentum
cutoff.  One might worry that radiative corrections in the
lattice theory could cancel this term so that instabilities
would not be cured.  However, the instabilities are associated
with large scalar field values, where the gauge symmetry is effectively broken.
Hence the runaway directions correspond to sectors of
the theory with weak gauge coupling, and so if the
sextic coupling $\mu_6$ appearing in \myref{irel6} is sufficiently large
the radiative corrections
cannot overpower the stabilizing term \myref{irel6}.

In fact, the flat directions imply that unbroken \NfourSYM\ is
not well behaved in the continuum limit at finite volume; the moduli are not fixed and the
partition function diverges because of the integral over the infinite
moduli space.  Therefore it will always be necessary to break \susy\
somehow.  In addition to adding \myref{irel6},
we advocate introducing antiperiodic boundary conditions
for the fermions in the temporal direction---finite temperature.
This lifts the moduli degeneracy in a way
that is removed by a zero temperature extrapolation,
such as working on an $L^3 \times 2L$ lattice, with
$2L/a$ sites in the temporal direction, and scaling
$L \to \infty$.  The antiperiodic
boundary conditions will also be beneficial to
numerical stability of the dynamical fermion
algorithms.

\subsubsection{Tuning with \susyc\ Ward identities}
If \susy\ is exact then the 
supercurrent $S_{\mu,i}$
is conserved.  The index $i$ corresponds
to $R$-symmetry, with the supercurrent transforming
as a {\bf 4} with respect to this group.
It follows from this supercurrent conservation
law that $\langle \partial^\mu S_{\mu, i}(x) \Ocal(y) \rangle$
vanishes at $x {\neq} y$ for all local operators $\Ocal$.    We can use this
property to fine-tune to a \susyc\ point in parameter space, a technique
pioneered in ${\cal N}=1$ \susy\ with Wilson fermions by the
DESY-M\"unster group \cite{Farchioni:2001wx}; here we discuss the extension
to \NfourSYM.

In the continuum, the supercurrent $S_{\mu ,i}$ is a linear combination of 
three dimension-7/2 operators.  It is easy to write down
corresponding lattice operators, though they not
unique.  In fact since we work with a cutoff theory, the lattice operators mix
with continuum operators of higher dimension 
in the same symmetry channel.  For this 
reason we express the operators $\Ocal_{\mu,i}$
in a continuum language, since our purpose is to convey the method
rather than the details of a specific implementation.

An analogous mixing analysis occurs in Wilson fermion lattice $\Ncal{=}1$ SYM.
The DESY-M\"unster group found that two dimension-7/2 
operators, the supercurrent $S_{\mu}$ and
another fermionic current $T_\mu$, mix in the lattice--continuum
matching.  For this reason it is necessary to combine two
corresponding lattice operators with
undetermined coefficients in order to find the
lattice operator that becomes $S_\mu$ in the continuum limit.

In our case we found in \cite{n4gw} that five dimension-7/2 operators
must be taken into account.  We denote them as
$\Ocal_{\mu, i}^{1\ldots 5}$, and the renormalized
supercurrent is in all generality of the form
\beq
&&
\begin{array}{l}\vspace{.05in}
\hspace{-.2in}S_{\mu,i}^{\text{ren.}}=
 \big\{ Z_1 \frac{1}{2}F_A{\cdot}\sigma\;\delta_{ij}
+ Z_2 \sqrt{2}\nott{D}\big(\phi_{ij}P_{\!L}{+}\phi^{ij} P_{\!R}\big)_A\\
\vspace{.05in}\hspace{.3in}
- Z_3 f^{ABC}\big(\phi^B_{ik}\phi^{kj}_C P_{\!R}+\phi^{ik}_B\phi^C_{kj}P_{\!L}\big)\big\}\gamma_\mu\psi_{\!jA} \\
\hspace{-.2in}+ \big\{ \!Z_4 \gamma_\nu F^A_{\mu\nu}\delta_{ij}
- Z_5 \sqrt{2}D_\mu \big(\phi^{ij}P_{\!L}+\phi_{ij}P_{\!R}\big)_{\!\!A}\!\big\} 
\psi_{\!jA} + \ord{a} \end{array}\nn\\&&
\hspace{.2in} \equiv  Z_n \Ocal^n_{\mu,i} + \ord{a}.
\label{scdf}
\eeq
The terms on the right-hand side are bare operators.
At tree level the supercurrent corresponds
to $Z_1{=}Z_2{=}Z_3{=}1$ and $Z_4{=}Z_5{=}0$.  The renormalization constants
$Z_{n}$ are universal with respect to the index $i$ due
to $SU(4)_R$ symmetry.

We wish to find the point in 
parameter space where the renormalized current
of \myref{scdf} satisfies $\p_\mu S_{\mu, i} = 0$ as an operator relation.
One therefore measures correlation functions containing $\p_\mu S_{\mu,i}$.
In actuality, we demand supercurrent conservation up to $\ord{a}$ corrections,
since at finite lattice spacing there will always be \susy-breaking
due to lattice artifacts.  What we seek is a trajectory in
parameter space such that \susy-breaking that
vanishes in the continuum limit.

Consider the SU(2) case, and suppose we have already tuned the
bare mass $m^2$ and the ratio of quartic couplings $\lambda_2/\lambda_1$
using the effective potential and susceptibility methods described
in \S\ref{sstff} above.  This leaves two more fine-tunings $Z_\phi$
and $\lambda_1$ to be performed using the \susyc\ Ward identities.
To tune these two parameters we need to examine correlation
functions of six operators
$\Ocal^n_{\mu, i}$ in the same symmetry channel as $S_{\mu, i}$. 
The natural choice is
the set $\Ocal^{1,\ldots 5}_{\mu,i}$ appearing
in \myref{scdf}, plus one dimension-9/2 operator
$\Ocal^6_{\mu,i}$.  One then measures the matrix of correlation functions
\beq
M^{mn}(t) \equiv \int d^3 {x}
\langle \Ocal^{m\dagger}_{0,i}(t,\mathbf{x}) \Ocal^n_{0,i}(0,\mathbf{0}) \rangle
\label{mdff}
\eeq
The $t$ derivative of this is the correlation function between $\p_\mu
\Ocal^m_{\mu,i}$ and $\Ocal^n_{0,i}$ at vanishing spatial momentum.

For the dimension-7/2 operators appearing in 
\myref{mdff}, correlation functions will
generically decay as
\beq
\bigvev{ \Ocal^{m\dagger}_{0,i}(t,\mathbf{x}) \Ocal^n_{0,i}(0,\mathbf{0}) }
\sim (t^2 + \mathbf{x}^2)^{-7/2}.
\eeq
Integrating with respect to $\mathbf{x}$, we therefore
find that these elements of $M^{mn}(t)$ to decay as 
\beq
M^{mn}(t) \sim t^{-4}.
\eeq
At the \susyc\ point and for the right choices of the coefficients $Z_m$ appearing
in \myref{scdf}, the corresponding combination
of correlation functions $Z_m M^{mn}$ will be suppressed
by the lattice spacing $a$ and hence decay as $a t^{-5}$ for the
$n$ associated with dimension-7/2 operators.  In fact,
GW fermions are automatically $\ord{a}$ improved, so
if the operators were likewise improved we could
even achieve a suppression $a^2 t^{-6}$ at the \susyc\
point, which would be easier to distinguish from the generic
$t^{-4}$ behavior.  Given the cost of the GW simulations,
and the fact that operator improvement is performed offline,
it would be well worth the effort.
For the dimension-9/2 operator we would require decays with $a t^{-6}$,
or $a^2 t^{-7}$ if improvement is performed.

These six conditions on the correlation matrix
$M^{mn}(t)$ fix the four ratios $Z_{2\ldots 5}/Z_1$
and the two parameters $Z_\phi, \lambda_1$ that must be
fine-tuned for the SU(2) and SU(3) cases.  In the SU($N_c >3$) case, additional
tunings with correlation functions of the supercurrent will be necessary.

\subsubsection{Other Ward identities}
\label{owid}
In the superconformal phase of $\Ncal=4$ SYM, $\vev{\phi_m}=0$,
the global symmetry of the theory is the supergroup $SU(2,2|4)$
and the R-symmetry group $SU(4)_R$.  The lattice preserves the
latter in a modified form, but deviations from the local (continuum) form
could perhaps serve as a measure of the lattice spacing.  The
supergroup $SU(2,2|4)$ includes conformal supercharges $\mathfrak{Q}_i$ other
than the four supercharges $Q_i$ corresponding to the supercurrents
$S_{\mu i}$ in \myref{scdf}.
The four additional supercurrents $\mathfrak{S}_{\mu i}$ could also be used
as a probe for the $\Ncal=4$ SYM theory, since they will have
their own Ward identities.  On the other hand, conservation
of $S_{\mu i}$ combined with scale invariance and
Poincar\'e symmetry implies conservation of $\mathfrak{S}_{\mu i}$,
so measurements of the additional Ward identities are
not independent, but rather serve as a means to check consistency with
predictions of the continuum theory.  Verification of this
feature would be reassuring in the regime of strong IR gauge coupling.

\subsubsection{Summary}
We have seen that for SU(2) and SU(3), there are four fine-tunings
in the action, $Z_\phi, m^2, \lambda_1, \lambda_2$. 
For SU($N_c >3)$ colors there are six,
$Z_\phi, m^2, \lambda_1, \ldots, \lambda_4$.  In addition, one
must fix the four relative renormalization constants $Z_{2}/Z_1, \ldots, 
Z_5/Z_1$ in the supercurrent.

It is conceivable that all but one of the 
scalar potential counterterms $\lambda_i$ can
be fixed by matching the effective potential to the \susyc\
scalar potential $\tr [\phi_m, \phi_n] [\phi_m, \phi_n]$,
though the practicality of this is yet to be established.
The overall strength of the quartic terms cannot be determined from the
effective potential.  The critical $m^2$ will be
determined from the effective potential.

Thus we see that in
the more optimistic scenario, where the effective potential
can be fully exploited, only two parameters of the lattice
action need to be fine-tuned by the \susyc\ Ward identities:
one fine-tuning of the bare kinetic coefficient $Z_\phi$ for the scalar,
one overall scalar potential coefficient $\lambda_1$, and
the four relative supercurrent coefficients $Z_1/Z_2, \ldots, Z_4/Z_5$.
Hence a total of six Ward identities must be measured.

If it proves too difficult to constrain all the ratios
$\lambda_{i+1}/\lambda_i$ of the quartic terms using effective
potential methods, then additional \susyc\ Ward identities
must be measured.  An intelligent strategy would be to
perform a combined minimization of all quantities that
can be measured with reasonable accuracy, so that in fact
the adjustment of parameters and mixing coefficients
is overconstrained.  Here, the additional Ward identities
mentioned in \S\ref{owid} could also be employed.

Aside from the challenges of developing an effective, automated
and optimized tuning strategy, one must also face the fact that
the lattice formulation employs GW fermions,
which are numerically expensive.  This is particullary
true in a theory such as this
one, with massless fermions and the corresponding critical slowing
down.

A first computation that 
needs to be done is to fix the multicanonical reweighting
function.  Small lattices ($4^4,6^4$) should suffice to get a rough idea
of how to proceed in further studies ($8^4,10^4,12^4$).  
Perturbative calculations may also help to narrow the
range of parameters that needs to be scanned, at least
for weaker couplings on smaller lattices.
Obviously early stages
of such work will be very much technical studies of the lattice
theory.   Nevertheless, we believe that
the beginnings of first principles nonperturbative study of
\Nfour\ SYM are not so far off.   As these progress,
it will be interesting to compare our results
to the on-going twisted \susy\ lattice simulation studies that
were initiated in \cite{Catterall:2008dv}.

\mys{Domain wall fermion lattice $\Ncal=1$ SYM}
\label{neq1}

\subsection{Domain wall fermions}
Lattice ${\cal N}=1$ super-Yang-Mills theory with
GW fermions requires no fine-tuning.
Domain wall fermions are a controllable approximation
to GW fermions, and we have recently performed large scale
simulations of the SU(2) theory \cite{Giedt:2008xm,Giedt:2008cd,
Gie08Latt}.
We  measured the gaugino condensate,
static potential, Creutz ratios and {\it residual mass} (a measure
of explicit chiral symmetry breaking arising from the domain wall 
approximation \cite{Blum:2000kn}).
With this data we extrapolated the gaugino
condensate to the chiral limit.  We review some
aspects of that study here.

\subsection{Introduction}
The only relevant or marginal operator allowed
in a gauge invariant lattice formulation of
pure $\Ncal=1$ super-Yang-Mills \cite{Ferrara:1974pu} (SYM) 
with hypercubic symmetry
is the gaugino mass term, as was emphasized long
ago in the analysis of \cite{Curci:1986sm}.  
As above, GW lattice chiral symmetry protects 
against additive renormalizations
of the gaugino mass in the continuum limit.  Hence the desired
continuum theory is obtained without fine-tuning of counterterms.

The domain wall fermion (DWF) that we use originates from
\cite{Kaplan:1992bt,Shamir:1993zy}.  Properly speaking, it is GW
only in the limit of infinite separation between the walls, $L_s \to \infty$.
In extrapolations this is often traded for the  residual
mass $\mres$, which is a {\it measure} of the explicit chiral
symmetry breaking.

Besides the absence of nonperturbative fine-tuning of the gaugino
mass, DWF have the advantage that the fermion
measure is positive and the square root of the determinant
which enforces the Majorana condition is analytic with a phase
that is independent of the gauge fields \cite{Neuberger:1997bg,Kaplan:1999jn}.
These three features are all lacking in the Wilson fermion formulation
that was applied in the only concerted lattice SYM effort to date,
by the \DMR\ collaboration \cite{Campos:1999du,Farchioni:2001wx,
Farchioni:2004ej,Farchioni:2004fy,Pee03,DMR_other}
and to a lesser extent Donini et al.~\cite{Donini:1997hh,Donini:1998pe,
Donini:1996nr}.  (Recently,
this program has been revived~\cite{DMR08}.)
Our research is a continuation of the work of
Fleming, Kogut and Vranas (FKV)~\cite{Fleming:2000fa} who first used
DWF for studying ${\cal N}=1$ SYM. Similar work has been
initiated by Endres \cite{End08a}, with an  extensive study
appearing recently \cite{Endres:2009yp}.
What sets the studies \cite{Giedt:2008xm,Endres:2009yp} 
apart is that an extensive scan of the
domain wall separation $L_s$ and measurement of
the residual chiral symmetry breaking mass $\mres$ was done at different values of the
bare lattice gauge coupling ($\beta=4/g^2=2.3$ and $2.4$ in our case) 
and spatial/temporal volumes ($L^3=8^3$ and $16^3$; $T=16,32$).
This has allowed for chiral extrapolations ($\mres \to 0$), and a preliminary view
on what occurs as we take the continuum, theormodynamic limit ($\beta,L,T \to \infty$).

It must be kept in mind that SYM does not have the Goldstone
phenomena and the lightest states are the analogues of $\eta'$ and
glueballs.  
The chiral symmetry that {\em is} broken
in SYM is a discrete symmetry, $Z_2$ in the SU(2) study that we review here.
The chiral regime is characterized by a
gaugino mass that is small compared to the particle
states of the theory, which should be of order $1/r_0$.  Here
$r_0$ is the  Sommer parameter \cite{Sommer:1993ce}, a measure
of the dynamically generated length scale
associated with confinement.  Some of
our results have $r_0\mres = 0.5$, which is 
too large, but other results have $r_0\mres=0.25$,
which ought to give chiral---hence supersymmetric---results to within 25\%.

\subsection{Lattice formulation}
The lattice formulation that is used in this study has already been
described by FKV~\cite{Fleming:2000fa,Giedt:2008xm}.
It employs Shamir DWF \cite{Shamir:1993zy} in the adjoint 
representation of SU(2) and the SU(2) fundamental plaquette Wilson gauge action.
The Majorana condition is imposed through a square root on the fermion
determinant, which as mentioned above is analytic and
introduces no gauge field dependent sign 
ambiguity \cite{Neuberger:1997bg,Kaplan:1999jn}.  

Lattice configurations were generated with a
dynamical fermion mass $m_f=0$, so that the finite size of the fifth dimension,
parameterized by $L_s$,
was the sole infrared regulator, through the corresponding
additive mass correction $\mres$, which is a measure of
residual chiral symmetry breaking \cite{Blum:2000kn}.

As was shown in the work by FKV, simulations 
performed at nonzero $m_f$, when extrapolated
to $m_f=0$, give identical results to the $m_f=0$ simulations.
Introducing $m_f$ requires more simulations due
to the $m_f \to 0$ extrapolation that must be done.
Provided $m_f=0$ simulations can be performed without crashing the
fermion inverter, we believe this is preferable. 

Added confidence in the $m_f=0$ simulations comes from comparing our
results to those of  \cite{Endres:2009yp}, as we discuss in
the next section.

\subsection{Bare gaugino condensate}
\label{bgco}
A summary of all results obtained in \cite{Giedt:2008xm} for the gaugino
condensate \condt\ is given in Tables
\ref{tall2p3} and \ref{tall2p4}.  

As mentioned above, confidence in the $m_f=0$ simulations comes from comparing our
results to those of Endres \cite{Endres:2009yp}.  He has performed
simulations with $m_f \not= 0$ and extrapolated to $m_f=0$, fitting
to the linear function $\cond a^3 = c_0 + c_1 m_f a$ at fixed $L_s$.  His
values of $c_0$ for $\beta=2.3$ on a $16^3 \times 32$ lattice
compare well to the condensate we report in Table \ref{tall2p3}
at the $L_s$ values that can be matched, $L_s=16, 24$; cf.~Table XII
of \cite{Endres:2009yp}.  Further comparisons will be discussed
below; cf.~Table \ref{endtab}.

Measurements were conducted on large and small lattice volumes;
it can be seen that in lattice units the finite-size dependence is mild
or insignificant for $\beta=2.3$ but quite noticeable for $\beta=2.4$.
This is sensible, given that $\beta=2.4$ corresponds to a finer lattice spacing,
and hence the physical volumes are smaller.  Results below will
show that the relative factor could be
as large as 2 (cf.~Table~\ref{stpr_2p3} vs.~Table~\ref{stpr_2p4}).

\begin{table}
\begin{center}
\begin{tabular}{|c|c|c|c|c|c|c|} \hline
$V \times T$   & $L_s$ & $\mres a$ & $\cond a^3$ & $\mres r_0$ & $\cond r_0^3$  \\ \hline
$8^3 \times 8$   & 16 & 0.158(5) & 0.00711(7) & --- & ---  \\ 
$8^3 \times 32$  & 16 & 0.181(3) & 0.00703(4) & 0.75(13) & 0.51(27) \\
$16^3 \times 32$ & 16 & 0.184(2) & 0.007051(5) & 0.668(10) & 0.337(11)  \\ 
\hline
$8^3 \times 32$  & 24 & 0.1541(15) & 0.005112(8) & 0.610(97)  & 0.32(15)  \\ 
$16^3 \times 32$ & 24 & 0.1564(17) & 0.005321(9) & 0.546(55)  & 0.226(68) \\ 
\hline
$8^3 \times 32$ & 32 & 0.1319(12) & 0.004321(11) & 0.501(69) & 0.24(10)  \\ 
$16^3 \times 32$ & 32 & 0.143(2) & 0.00445(2) & 0.483(58) & 0.172(61)  \\ 
\hline
$8^3 \times 32$ & 40(I) & 0.1183(54) & 0.00383(3) & --- & ---  \\ \hline 
$8^3 \times 16$  & 48 & 0.1043(17) & 0.003563(20) & 0.361(31) & 0.148(37) \\ 
$8^3 \times 32$  & 48 & 0.1071(10) & 0.003551(11) & 0.409(31) & 0.198(45)  \\ \hline 
$8^3 \times 32$  & 64 & 0.08864(84) & 0.003164(10) & 0.300(35) & 0.122(42) \\ 
\hline
\end{tabular}
\caption{The gaugino condensate \condt\ and residual mass $\mres$ for various lattice 
sizes and $L_s$ values, all at $\beta=2.3$.
The $L_s=40$ value, with an ``(I)'' after it, is obtained by the
interpolation method described in the text.  
Values in units of the Sommer parameter $r_0$ are also shown, for those cases where the
potential was measured (in particular, for all points that are included in the
chiral extrapolation fit).  The $L_s=16$ data was not included
in the linear chiral extrapolation fit, because these points had
too much curvature (with respect to $\mres a$) associated with them.
\label{tall2p3} }
\end{center}
\end{table}

\begin{table}
\begin{center}
\begin{tabular}{|c|c|c|c|c|c|} \hline
$V \times T$ & $L_s$ & $\mres a$ & $\cond a^3$ & $\mres r_0$ & $\cond r_0^3$ \\ \hline
$8^3 \times 32$  & 16 & 0.080(2)  & 0.004839(15) & 0.547(30) & 1.55(22) \\
$16^3 \times 32$ & 16 & 0.0969(8) & 0.00499(6)   & 0.5355(66) & 0.842(25) \\
\hline
$8^3 \times 32$  & 24 & 0.0601(15) & 0.003293(17) & 0.417(26) & 1.10(18) \\  
$16^3 \times 32$ & 24 & 0.0838(17) & 0.00389(8) & 0.385(35) & 0.38(10) \\
\hline
$16^3 \times 32$ & 28(I) & 0.0721(33)  & 0.003452(45)  & --- & ---  \\ %
\hline
$8^3 \times 32$  & 32 & 0.0486(12) & 0.00269(2) & 0.296(15) & 0.61(08)  \\ 
$16^3 \times 32$ & 32 & 0.0653(15) & 0.003330(12) & 0.313(33) & 0.37(11) \\ 
\hline
$8^3 \times 32$  & 40(I) & 0.0390(24) & 0.00234(8) & --- & --- \\ % 
\hline
$8^3 \times 32$  & 48 & 0.0328(9) & 0.002165(18) & 0.224(17) & 0.69(15) \\  
\hline
\end{tabular}
\caption{Results similar to Table \ref{tall2p3},
except that these are for $\beta=2.4$.
\label{tall2p4} }
\end{center}
\end{table}

We measured the condensate at other values of $L_s$ using
a sea-$L_s$/valence-$L_s$ approach.  The condensate was
measured using DWF with $L_s^{\text{val.}}$ on top of dynamical lattices produced
using a nearby $L_s^{\text{sea}}$.  Performing this for $L_s^{\text{sea}}$
values on either side of $L_s^{\text{val.}}$ yields robust interpolated (I)
results.

We also used the results of the static potential
study summarized in \S\ref{stpo} below to express
$\mres$ and $\cond$ in terms of the Sommer scale $r_0$.
Note that the $\beta=2.4$ value of $\mres r_0$ at $L_s=48$
indicates that the effective gaugino mass (which should
be approximately equal to $\mres$) is roughly 1/4 the inverse
Sommer scale, so that we are beginning to enter
the chiral regime where \susy\ is well approximated.
On the other hand, it can be seen that $\mres r_0$ is
unpleasantly large for $\beta=2.3$ with $L_s \leq 32$, and likewise
the condensate in physical units is small compared to
the $\beta=2.4$ results.  Clearly $\beta=2.3$ is further away from the 
supersymmetric limit due to the coarser lattice.  On the other hand
it can be seen that the $\beta=2.4$ data shows a marked
volume dependence due to the smaller physical ``box'' that
the states must squeeze into.

\subsubsection{Combined results for $\beta=2.3$}
Here we compar in detail our results
to those of Endres \cite{Endres:2009yp}.
At each $L_s$ he has fit $m_f \not=0$ results to
the form:
\beq
\cond a^3 = c_0 + c_1 ~ m_f a .
\label{mfeq}
\eeq
What is interesting about the results of his Table XI is
that the coefficient $c_1$ shows a regular pattern with
respect to $L_s$:
\beq
c_1(L_s) = 0.0828(1) + 0.00025(7) \times (L_s - 28).
\label{c1eq}
\eeq
Here, the numbers in parentheses represent our estimate
of the error in this formula, based on his results.  From
this formula we can use his measurements of $\cond a^3$ at $m_f=0.02$
for $L_s=32, 40, 48$ to obtain predictions of the $m_f=0$
condensate.  These are shown, combined with his extrapolation 
values at $L_s=16, 20, 24, 28$ in Table~\ref{endtab}.  Comparing
to our Table~\ref{tall2p3}, results of which we reproduce
in the two right-most columns of Table \ref{endtab},
we find that there is reasonable consistency.
We will use these combined results in the chiral
extrapolation fit below.

\begin{table}
\begin{center}
\begin{tabular}{|c|c|c|c|} \hline
 & Endres/Extrap. & Us ($8^3 \times 16$) & Us ($16^3 \times 32$) \\
$L_s$ & $\cond a^3$ ($16^3 \times 32$) & $\cond a^3$  & $\cond a^3$ \\ \hline
16 & 0.0070544(51) & 0.00703(4) & 0.007051(5)  \\
20 & 0.0058979(55) & --- & --- \\
24 & 0.0051697(49) & 0.005112(8) & 0.005321(9) \\
28 & 0.0046770(51) & --- & --- \\ \hline
32 & 0.00432(2) & 0.004321(11) & 0.00445(2) \\
40 & 0.00381(2) & 0.00383(3) & ---  \\
48 & 0.00346(3) & 0.003551(11) & --- \\ \hline
\end{tabular}
\caption{Column 2 contains extrapolation results of Endres for $L_s=16,20,24,28$ with
$\beta=2.3$ on the $16^3 \times 32$ lattice.  Also in this column
are the results we obtain for $L_s=32,40,48$
from his data.  Here, we model the behavior of $c_1$ from \myref{mfeq} as given by
\myref{c1eq}, extracted from his results.  We then use 
his $m_f=0.02$ results at these $L_s$ values to estimate
the $m_f=0$ condensate.  In columns 3 and 4 we give
our simulation results for comparison.  \label{endtab} }
\end{center}
\end{table}

\subsection{Gluonic observables}
\label{gluob}
\subsubsection{Static potential}
\label{stpo}
The static potential was obtained by measuring Wilson
loops with one side of length $t$ in the temporal direction, according
to standard methods.  Having obtained $V(r) a$ from fitting
the exponential decay in time, we fit the data to the standard form
\beq
V(r) a = V_0 a + \s a^2 (r/a) - \frac{\alpha}{r/a}.
\eeq
We obtain the Sommer parameter $r_0/a$ from this fit, using the formula
\beq
\frac{r_0}{a} = \sqrt{\frac{1.65-\alpha}{\s a^2}}.
\eeq
This approach to the determination of $r_0/a$ has
some sensitivity to the range of radii that is fit,
and obviously depends on what form we assume for $V(r) a$.

The results of our static potential fit are
presented in Tables~\ref{stpr_2p3} and \ref{stpr_2p4}.
For the $L=16$ results,
the fits were also done using the same set of Wilson loops as
in the $L=8$ case, denoted ``$L=8$ method'', so that dependence
choice of Wilson loops could be controlled
for, and therefore ruled out as a spurious source of finite
size dependence.

These results are to be compared with Table 1 of \cite{End08a}
or Table IX of \cite{Endres:2009yp}.
There, a nonzero fermion mass was used
different fit ranges for $r,t$ were employed.
In particular, our fits include the
points $r/a=1,\sqrt{2}$, which have very small errors and
thus strongly influence the fit.  Thus for $L_s=16$
we have also performed a fit with $r_{\min} = \sqrt{3}$ as 
was done in \cite{End08a}, as can be seen in the second $L_s=16$
entry for $\beta=2.3,2.4$ in Tables \ref{stpr_2p3} and \ref{stpr_2p4}.  
Our results for the fit quantities at $\beta=2.3$ are
in good agreement once this restriction is imposed.
For our other $L_s$ values we have far fewer samples,
as larger $L_s$ comes at greater computing cost.  The degradation
of statistical errors that results if we exclude the
$r=1,\sqrt{2}$ points is unacceptable, which is why we do
not quote results with the same $r_{\min} = \sqrt{3}$ as \cite{End08a}
for the other $L_s$ values.  On the other hand, it can be
seen from the $L_s=16$ results that the choice of $r_{\min}$
only has a 10\% effect on the $r_0/a$ estimate, so that the
choice of $r_{\min}$ is not crucial to the broad picture that
we are after.
The $\beta=2.4$ results are also in reasonable
agreement with \cite{End08a}, comparing to the numbers
we obtain at $16^3 \times 32$, $L_s=16$, with $r_{\min}=\sqrt{3}$,
and keeping in mind the nonzero $m_f$ in \cite{End08a}.

\begin{table}
\begin{center}
{\footnotesize
\begin{tabular}{|c|c|c|c|c|c|c|c|}  \hline
$V \times T$ & $L_s$ & $V_0 a$ & $\s a^2$ & $\alpha$ & $r_0/a$  & $\s r_0^2$ & method \\ 
\hline
$8^3 \times 32$ & 16 & 0.717(83) & 0.074(28) & 0.368(55) & 4.16(73) & 1.282(55) & $L=8$ \\
$16^3 \times 32$ & 16 & 0.6533(80) & 0.1004(25) & 0.3271(57) & 3.630(39) & 1.3229(57) & $L=16$ \\
$16^3 \times 32$ & 16 & 0.489(45) & 0.1367(88) & 0.159(55) & 3.303(50) & 1.491(55) & $L=16,r\geq\sqrt{3}$ \\ 
\hline
$8^3 \times 32$  & 24 & 0.718(89) & 0.082(30) & 0.371(60) & 3.96(63) & 1.279(60) & $L=8$ \\
$16^3 \times 32$ & 24 & 0.752(70) & 0.102(24) & 0.411(46) & 3.49(35) & 1.239(46) & $L=16$ \\
$16^3 \times 32$ & 24 & 0.696(67) & 0.119(22) & 0.372(46) & 3.27(24) & 1.278(46) & $L=8$ \\
\hline
$8^3 \times 32$  & 32 & 0.748(82) & 0.087(27) & 0.400(55) & 3.80(52) & 1.250(55) & $L=8$ \\
$16^3 \times 32$ & 32 & 0.745(90) & 0.109(31) & 0.412(59) & 3.38(40) & 1.238(59) & $L=16$ \\
$16^3 \times 32$ & 32 & 0.635(50) & 0.146(17) & 0.338(33) & 3.00(14) & 1.312(33) & $L=8$ \\
\hline
$8^3 \times 16$  & 48 & 0.706(68) & 0.107(22) & 0.372(47) & 3.46(29) & 1.278(47) & $L=8$ \\
$8^3 \times 32$  & 48 & 0.768(47) & 0.085(15) & 0.414(33) & 3.82(29) & 1.236(33) & $L=8$ \\
\hline
$8^3 \times 32$  & 64 & 0.680(94) & 0.113(32) & 0.353(63) & 3.38(39) & 1.297(63) & $L=8$ \\
\hline
\end{tabular}
}
\caption{Gluonic observables obtained from the static potential for $\beta=2.3$.
The quantity $r_0$ is physical, so that the ratio $r_0/a$ provides
a measure of the lattice spacing.
\label{stpr_2p3}
}
\end{center}
\end{table}

\begin{table}
\begin{center}
{\footnotesize
\begin{tabular}{|c|c|c|c|c|c|c|c|}  \hline
$V \times T$ & $L_s$ & $V_0 a$ & $\s a^2$ & $\alpha$ & $r_0/a$  & $\s r_0^2$ & method \\ \hline
$8^3 \times 32$ & 16 & 0.617(11) & 0.0292(30) & 0.2857(83) & 6.84(33) & 1.3643(83) & $L=8$ \\
$16^3 \times 32$ & 16 & 0.5846(32) & 0.04531(91) & 0.2659(24) & 5.526(51) & 1.3841(24) & $L=16$ \\
$16^3 \times 32$ & 16 & 0.537(11)  & 0.0554(20)  & 0.219(15)  & 5.083(67) & 1.431(15) & $L=16, r \geq \sqrt{3}$ \\
\hline
$8^3 \times 32$ & 24 & 0.636(12) & 0.0280(33) & 0.2997(88) & 6.94(39) & 1.3503(88) & $L=8$ \\
$16^3 \times 32$ & 24 & 0.579(40) & 0.065(13) & 0.272(27) & 4.60(41) & 1.378(27) & $L=16$ \\
\hline
$8^3 \times 32$  & 32 & 0.609(12) & 0.0369(36) & 0.2809(90) & 6.09(28) & 1.369(90) & $L=8$ \\
$16^3 \times 32$ & 32 & 0.611(43) & 0.059(13)  & 0.295(29)  & 4.79(50) & 1.355(29) & $L=16$ \\
\hline
$8^3 \times 32$  & 48 & 0.648(15) & 0.0288(44) & 0.309(11) & 6.83(49) & 1.341(11) & $L=8$ \\
\hline
\end{tabular}
}
\caption{Gluonic observables obtained from the static potential for $\beta=2.4$.
We note that $r_0/a$ shows significant volume dependence for $L_s=32$.  Comparing
to Table \ref{stpr_2p3}, the $L,L_s \to \infty$ trend seems to indicate
a lattice spacing that is slightly more 1/2 of the $\beta=2.3$ value,
in agreement with what we estimate using two-loop renormalization
of the gauge coupling.
\label{stpr_2p4}}
\end{center}
\end{table}

Above, we have used the results of Tables \ref{stpr_2p3} and \ref{stpr_2p4}
to scale the residual mass and condensate
to $r_0$ units.  (Note that the $r_0/a$ values with identical
lattice parameters were used in this procedure,
rather than a uniform $r_0/a$ value across all $\mres$ and $\cond$.)
With the string tension in hand, we now see that the energy
scale of confinement $\sqrt{\s r_0^2} \approx 1.4$ lies above the explicit chiral symmetry
breaking scale $\mres r_0$ by a factor of 1.8 to 3.8 for $\beta=2.3$, and 3.0 to 5.2
for $\beta=2.4$.  This is consistent with the observation that the
string tension results in Tables \ref{stpr_2p3} 
and \ref{stpr_2p4} are insensitive to the range of
$L_s$ values displayed there, when expressed in physical units ($\s r_0^2$).
That is, confinement dynamics are to
a good approximation decoupled
from the explicit chiral symmetry breaking.  Since the lowest lying excitations
of SYM are glueballs and superpartners, the gap associated with confinement
should also decouple these states from the explicit chiral symmetry
breaking.  Thus it appears that we in the regime where the spectrum
reflects supersymmetry, modulo lattice artifacts, 
and it will be quite interesting to examine the
spectrum in order to check whether or not this is true---something
we will do in future work.

At this point we make some further remarks regarding
finite size effects, since the volumes are for several
points small.  Because there are
no Goldstone modes the situation is very different from QCD,
where one has to watch out for the pion Compton wavelength.
Instead, here we have to watch out for the $\eta'$ 
Compton wavelength, which will be of order $r_0$.  

For the
$16^3 \times 32$ lattices at $\beta=2.3$, we have $r_0/a =$ 3.0 to 3.5, which
gives a lattice extension 4.6 $r_0$ to 5.3 $r_0$.  If the
glueballs and $\eta'$ have wavelength $r_0$, we should be
safe from finite volume effects.  On the $8^3$ lattices
there is a bigger problem, but still the lattice extent
is always at least $2 r_0$ for the points that are used
in our chiral extrapolation for $\beta=2.3$.
In fact, our results in Table II show only a
mild volume dependence for bare lattice quantities,
confirming the arguments we have just made.

The volume dependence for $\beta=2.4$ is more
significant, as we have discussed above. For
the $16^3 \times 32$ lattices we have lattice extent $3.3 r_0$ to $3.5 r_0$
in space-like directions.
For the $8^3$ lattices we have lattice extent $1.2 r_0$,
which is too small.  This likely explains
why the $L=8$ versusy $L=16$ results are so
different for the measured quantites.  In fact, it may
only be the $L=8$ results that are significantly impacted
by finite size effects; the $L=16$ 
lattices, having $3.3 r_0$ to $3.5 r_0$, are perhaps
large enough.  Evidence in that direction is given by
the insensitivity of the $\beta=2.3$ results
for lattices of this size, in units of $r_0$.

In the case of $\beta=2.4$, further simulations should eventually
be done on lattices larger than $16^3 \times 32$, but it is a major undertaking
that is beyond our present capabilities, which are at
present state of the art.  

\subsection{Extrapolation of the gaugino condensate}
\label{extr}
In our previous work, we presented only a linear extrapolation
of the condensate in $\mres$, to be discussed in \S\ref{linet}.
Here we will also present some new results where we
perform nonlinear chiral extrapolations based on the functional
forms that Endres has assumed in \cite{Endres:2009yp}.  It will
be seen that although our simulations are in agreement,
different assumptions about the functional form of
the chiral extrapolation lead to $\mres \to 0$ estimates
that differ by factors as large as 3, as has already been
noted in \cite{Endres:2009yp}.

\subsubsection{Linear extrapolation}
\label{linet}
\begin{figure}
\begin{center}
\includegraphics[width=3in,height=5in,angle=90]{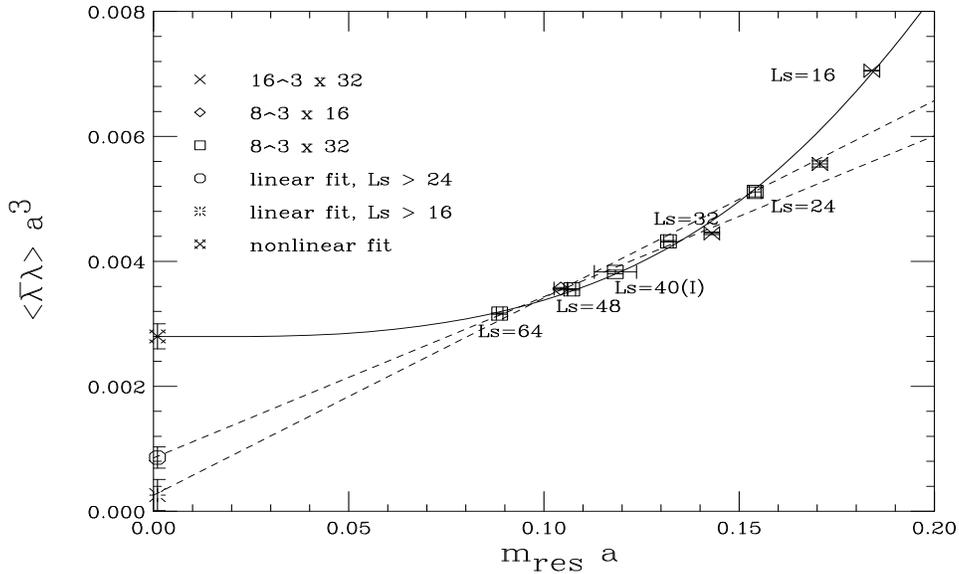}
\caption{Condensate vs.~\mrest\ for $\beta=2.3$, in bare lattice
units.  Dashed lines show the two linear fits (differing
by the minimum $L_s$ included).  Extrapolated values
together with fit errors are shown at $\mres=0$. 
The Solid line combines results from fitting 
functions \myref{endfcn} and \myref{mrfit}.
\label{cvm2p3} }
\end{center}
\end{figure}

\begin{figure}
\begin{center}
\includegraphics[width=3in,height=5in,angle=90]{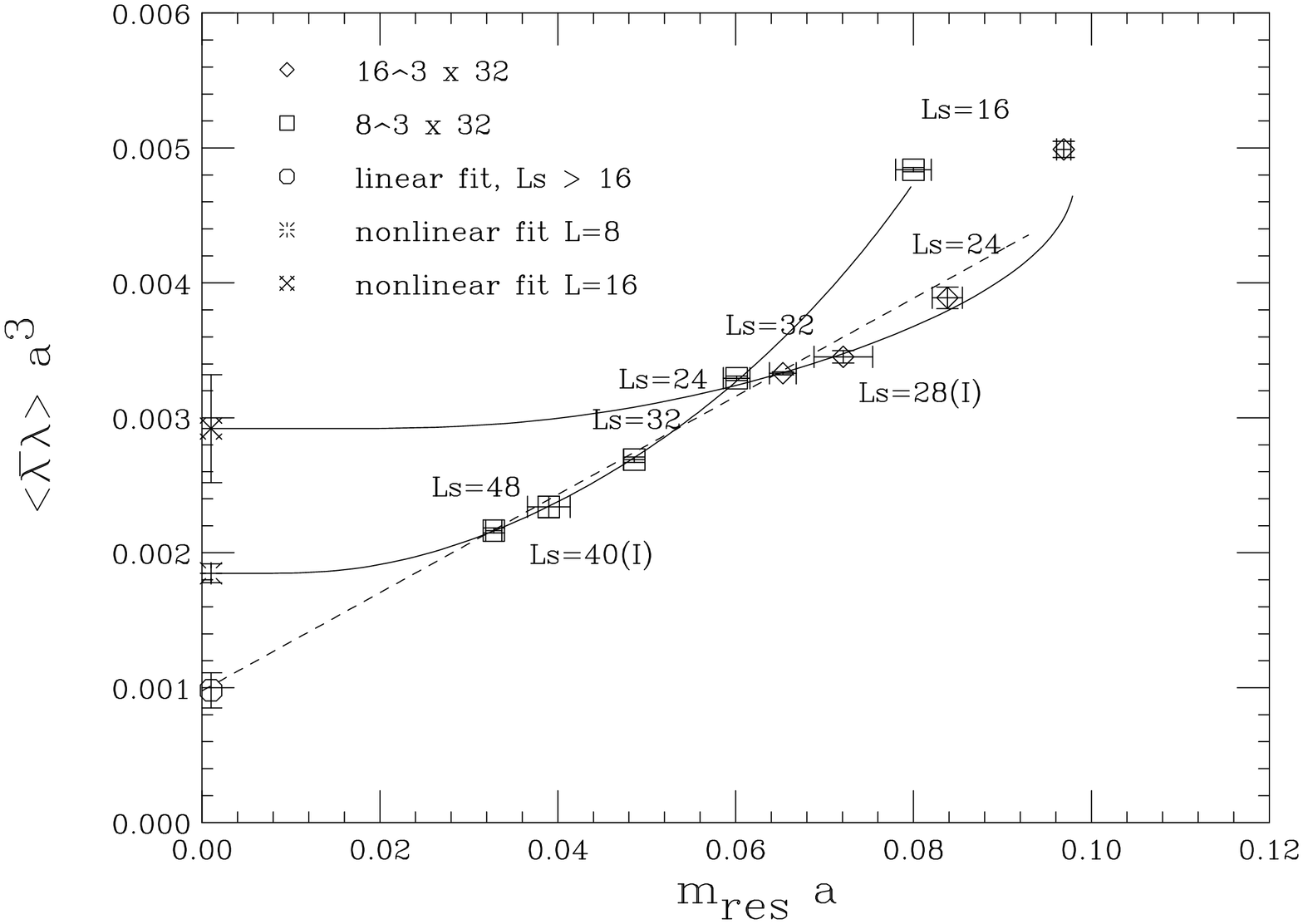}
\caption{Condensate vs.~\mrest\ for $\beta=2.4$, in bare lattice
units.  The dashed line shows the linear fit.  The extrapolated
value together with fit error is shown at $\mres=0$. 
Solid lines combine results from fitting 
functions \myref{endfcn} and \myref{mrfit}.
\label{cvm2p4} }
\end{center}
\end{figure}

One important question is the size of $L_s$ necessary to get into the linear
regime where 
\beq
\cond a^3 = c_0 + c_1 \mres a
\label{conex}
\eeq
is a good approximation.  Obviously, this serves as an indicator of
where we need to be in order to have SYM well-approximated.  Thus,
the measurement of \condt\ vs.~\mrest\ is an important benchmark
for determining the regime in which other SYM phenomena can be studied
with the DWF lattice approach.
Another question is the extent to which $c_{0,1}$ are sensitive to
finite spacetime volume ($V_4=V \times T = L^3 \times T$ in our notation). 
Interestingly, we find that most of the volume dependence is absorbed into \mrest,
as can be seen from the dashed lines
in Figs.~\ref{cvm2p3} and \ref{cvm2p4}.  
To a good approximation the $8^3 \times 32$ and $16^3 \times 32$
lattice data lie on the same line.  The smaller value of \mrest\ on
the smaller lattice is most likely due to a smaller density of
near-zero modes.  The linear chiral extrapolation ($\mres \to 0$) of $\cond a^3$
obtained from the fit to \myref{conex} is given in Table \ref{tbex}.  A feel
for the sensitivity to the fitted range of $L_s$ can be
seen from the two results we provide for $\beta=2.3$,
which differ by the minimum $L_s$ that was included.  
In fact, the quality of the $L_s>16$ fit is very poor
due to nonlinear dependence on $\mres$ that enters at $L_s=24$,
as can also be seen from Fig.~\ref{cvm2p3}.

\begin{table}
\begin{center}
\begin{tabular}{|c|c|c|c|c|} \hline
$\beta$ & $L_s$ range & $\chi^2/d.o.f.$ & $c_0$ & $c_1$ \\ \hline
2.3 & 24-64 & 136 & 0.00026(25) & 0.0316(20) \\
2.3 & 32-64 & 30.0 & 0.00086(17) & 0.0258(15) \\ \hline
2.4 & 24-48 & 19.5 & 0.00098(13) & 0.0364(23) \\ 
\hline
\end{tabular}
\caption{Fit results for the linear extrapolation \myref{conex}
of the gaugino condensate in $\mres$, depending upon the range of $L_s$ values used.
For $\beta=2.3$, the quality of the $L_s>16$ fit is very poor
due to nonlinear dependence on $\mres$ that enters at $L_s=24$,
as can also be seen from Fig.~\ref{cvm2p3}.
\label{tbex}}
\end{center}
\end{table}

\subsubsection{Nonlinear extrapolation}
We now follow Endres \cite{Endres:2009yp} and fit the data to the form
\beq
\cond a^3 = b_0 + \frac{b_1}{L_s} \exp ( -b_2 L_s) .
\label{endfcn}
\eeq
All $\beta=2.3$ results of Tables \ref{tall2p3} and \ref{endtab}
except our $L_s=16$ results for $8^3 \times 8$ and $8^3 \times 32$ lattices
are included.  We then exclude $L_s=16$ data, and then
$L_s=16,20$ data.  The results of the fit are presented
in Table \ref{bigext}.  The chiral limit of $\cond a^3$
is given by $b_0$.  It can be seen that $\cond a^3 = 0.0027(2)$
is a good characterization of the result.  We note that
this is much larger than the value obtained by linear extrapolation
in $\mres$.

In Table \ref{bigext2} we fit our $\beta=2.4$ to the same
formula.  Due to the large finite size effects at this
finer lattice spacing, we have separately fit the $L=8$ and
$L=16$ data.  The dependence on $L$ is also strong
in the chiral extrapolation of the condensate, $b_0$.  We
would need an additional $L$ value, such as $L=12$, in order
to obtain an estimate of the condensate in the $L \to \infty$
limit.  Nevertheless we see that the trend is for the condensate
to increase with $L$, and we obtain an estimate of
$\cond a^3 \gappeq 0.0029$ in the chiral limit. We argued
above that the $L=16$  lattices, having $3.3 r_0$ to $3.5 r_0$, 
are perhaps large enough to give a good approximation
of the $L \to \infty$ behavior.  If that is true,
then we obtain an estimate for the $\beta=2.4$
chiral limit of $\cond a^3 = 0.0029(4)$.  Note
that this is significantly larger than the
result of linear extrapolation in $\mres$.

\begin{table}
\begin{center}
\begin{tabular}{|c|c|c|c|c|} \hline
$L_s$ range & $\chi^2/d.o.f.$ & $b_0$ & $b_1$ & $b_2$ \\ \hline
16-64 & 24.5 & 0.002796(96) & 0.0950(24) & 0.0209(27) \\
20-64 & 26.6 & 0.00270(16)  & 0.0901(53) & 0.0171(51) \\
24-64 & 28.6 &	0.00275(20)  &	0.093(10)  & 0.0193(78) \\ \hline
\end{tabular}
\caption{Fit of equation \myref{endfcn} for $\beta=2.3$.
\label{bigext} }
\end{center}
\end{table}

\begin{table}
\begin{center}
\begin{tabular}{|c|c|c|c|c|} \hline
$L$ & $\chi^2/d.o.f.$ & $b_0$ & $b_1$ & $b_2$ \\ \hline
8  & 0.699 & 0.001848(71) & 0.0792(82) & 0.0342(62) \\
16 & 3.27 & 0.00292(40) & 0.084(51) & 0.058(49) \\ \hline
\end{tabular}
\caption{Fit of equation \myref{endfcn} for $\beta=2.4$,
using all data from Table \ref{tall2p4} for each $L=8,16$.
\label{bigext2} }
\end{center}
\end{table}

Endres has argued that the residual mass should be fit to
\beq
\mres a = \frac{1}{L_s} ( a_0 \exp(-a_1 L_s) + a_2),
\label{mrfit}
\eeq
a functional form that was derived by transfer
matrix methods in Eq.~(3.7) of \cite{Christ:2005xh}.
Endres has shown that \myref{mrfit} is consistent with his $m_f=0.02$ data.
Here we have performed a fit to this function using
our $m_f=0$ data.  The results
are given in Table \ref{mrestab}.  In order to
obtain stable fits, it was necessary to
separate the $\mres$ data for $\beta=2.4$ according the
two $L$ values.  One can see that the finite size
dependence of the fit parameters is significant in
that case.

\begin{table}
\begin{center}
\begin{tabular}{|c|c|c|c|c|c|c|} \hline
$\beta$ & $L$ & $L_s$ range & $\chi^2/d.o.f.$ & $ a_0$ & $a_1$ & $a_2$ \\ \hline
2.3 &	all	&16-64&	4.47&	-5.71(22) & 	0.0308(58) & 	6.44(38) \\ \hline
2.4 & 	8&	16-48&	0.250&	-1.64(56) & 	0.102(24) & 	1.596(28) \\
2.4 &	16 & 	16-32 & 	0.237 & 	-20(17) &	0.226(55) & 2.097(34) \\ \hline
\end{tabular}
\caption{Fit of $\mres$ data to \myref{mrfit}. \label{mrestab} }
\end{center}
\end{table}

Combining the results of the fit functions \myref{endfcn} and \myref{mrfit},
we obtain a map $L_s \to (\mres a,\cond a^3)$ for arbitrary $L_s$.
With this we obtain the nonlinear fit curves that appear in Figs.~\ref{cvm2p3}
and \ref{cvm2p4} as solid lines.  

\subsubsection{Discussion}
The message  is that one needs
data at smaller $\mres$. The nonlinear
fits are motivated by known features of the DWF discretization
and are more likely to be trustworthy at the large $\mres$ values
that we are working with.  Still, it would be greatly reassuring
to have simulations at small enough $\mres$ that the
linear extrapolation agrees with these formulae.  At
a minimum, simulations at larger $L_s$ should be performed
as a test of the predictions made by the solid curves 
in Figs.~\ref{cvm2p3} and \ref{cvm2p4}.

According to Table \ref{stpr_2p4}, the value of the lattice spacing $a$ is smaller
on the $\beta=2.4$ lattice.
Thus it is surprising that the extrapolated values of $\cond a^3$ is as
large for $\beta=2.4$ as it is for $\beta=2.3$.    
A plausible interpretion is that there are larger renormalizations 
of the condensate on a finer lattice ($\beta=2.4$),
a hypothesis that we are preparing to test with
nonperturbative renormalization \cite{Martinelli:1994ty,Blum:2001sr} in an upcoming study.

\subsection{Outlook}
\label{conc}
Future work that is envisioned aims to
develop a deeper understanding of the configurations
that are responsible for generating the nonzero gaugino condensate.
In particular, we would like to elucidate the continuum
picture on the cylinder $\Rbf^3 \times S^1$,
where it is monopoles and ``KK monopoles'' that
combine to yield the infinite volume value \cite{Davies:1999uw}.
It is already known from spectral flow techniques
that fractional topological charge plays an important
role in the zeromodes of adjoint lattice fermions \cite{Edwards:1998dj},
and we would like to explicitly connect that with the semiclassical
configurations that have been suggested by continuum
methods.  It would also be interesting to study questions
of {\it center dominance} and {\it abelian dominance} in LSYM,
following the line of research that has been pursued
in the ``pure glue'' (i.e.~quenched) SU(2) theory \cite{Ambjorn:1999gb,
Ambjorn:1999ym,DelDebbio:1998uu,Giedt:1996zp}.

Note that rather large $L_s$ values were required
in order to get $\mres r_0 \sim 1/4$.  To improve the situation
we envision switching to simulations with modified versions of DWF
that have superior chiral behavior \cite{Vranas:2006zk,Brower:2004xi}.
This should clear up uncertainties in the chiral extrapolation
that are apparent from the results in \S\ref{extr} above.

\mys{Conclusions}
As algorithms and hardware have advanced, so have the
prospects for simulating four-dimensional \susyc\ gauge
theories.  We have illustrated in three different types
of models how GW chiral symmetry and exact lattice flavor
symmetries can be used to limit the counterterms that
must be adjusted to get the desired continuum limit.
Furthermore, it was seen that in most cases only bosonic
counterterms need to be adjusted.  Multicanonical reweighting
techniques provide a viable method for accomplishing this
on small lattices in the near term.  On the other hand,
the $\Ncal=1$ SYM study that we have concluded with illustrates
the magnitude of the challenge in simulating larger lattices.
In fact, that study consumed 30 million IBM BlueGene/L node
hours.  Nevertheless, continuing progress in the program of
simulating \susyc\ gauge theories is to be expected and
we look forward to reporting our advances in this field
as further results, especially in SQCD and $\Ncal=4$ SYM,
are obtained.

\section*{Acknowledgements}
We express our thanks to collaborators on the research
reviewed above:  Richard Brower, Simon Catterall, Joshua Elliot,
George Fleming, Guy Moore, Pavlos Vranas.
We received support from Rensselaer faculty development funds.

\myappendix

\mys{Super-QCD continuum theory}
\label{sqcth}
Here we describe the supersymmetric continuum
theory with massless ``quarks'' and scalar partners, ``squarks.''
Since lattice studies of SQCD are
intended for understanding new strong interactions
that are associated with spontaneous \susy\ breaking
and its transmission to the visible sector, these
terms are only used by way of analogy; the $SU(N)$ gauge
symmetry is in addition to the Standard Model gauge
group, and is strongly coupled at scales of a TeV or
perhaps much more.

The Minkowski space Lagrangian is
\beq
\Lcal & =& -\half \tr G_{\mu \nu} G^{\mu \nu} - 2i \tr \lambar \sbar^\mu \Dcal_\mu \lambda
+ \tr D^2 - (\Dcal_\mu p)^\dagger \Dcal^\mu p \ddd -i \chib_p^T \sbar^\mu \Dcal_\mu \chi_p
+ F_p^\dagger F_p + i \sqtw g p^\dagger \lambda \chi_p
-i \sqtw g \chib_p^T \lambar p \ddd + g p^\dagger D p
+ (p \to q).
\label{csqcd}
\eeq
Here we mostly follow the two-component conventions of \cite{Wess:1992cp}, except that
we use $\mu,\nu$ for spacetime indices and write the ``gluon''
field strength $G_{\mu \nu}$.  We have auxiliary
scalar fields $D=D^a t^a$ and $F_p$ which allow for \susy\
without imposing the equations of motion (off-shell \susy).
They are easily integrated out using their equations
of motion.
Flavor indices have been suppressed, and our convention
for the generators is $\tr t^a t^b = (1/2) \delta^{ab}$.
As in \S\ref{sqcd}, we use a matrix notation in SU(N) color space,
so that $\chib_p^T$ is supposed to represent an $N$-component row vector.
Covariant derivatives $\Dcal_\mu$
take the usual form.  The Yukawa terms can be related to those
of the main text by the redefinition
\beq
\lambda \to i \lambda, \quad \lambar \to -i \lambar.
\eeq
Once this is done, it is clear that the CP and $S$ symmetries
that have been imposed in \S\ref{sqcd} are also symmetries of
the continuum theory \myref{csqcd}.
The \susy\ transformations under which \myref{csqcd}
is invariant are, for the matter fields,
\beq
&&\delta_\xi p = \sqtw \xi \chi_p, \quad
\delta_\xi \chi_p = i \sqtw \s^\mu \xibar \Dcal_\mu p + \sqtw \xi F_p,
\ddd \delta F_p = i \sqtw \xibar \sbar^\mu \Dcal_\mu \psi
+ i 2 g \xibar \lambar p, \quad (p \to q),
\eeq
and for the gauge fields,
\beq
&&\delta_\xi A_\mu = -i \lambar \sbar_\mu \xi + i \xibar \sbar_\mu \lambda,
\quad \delta_\xi \lambda = \s^{\mu \nu} \xi G_{\mu \nu} + i \xi D,
\ddd \delta_\xi D = -\xi \s^\mu \Dcal_\mu \lambar - \Dcal_\mu \lambda \s^\mu \xibar,
\eeq
where $\xi$ is a Grassmann transformation parameter.
It is straightforward to continue to imaginary
time and the Euclidean version of the action that
we use in \S\ref{sqcd}.

\bibliography{rvw08}
\bibliographystyle{JOEL}
\end{document}